\documentclass{aa}  

\usepackage{hyperref}
\usepackage{graphicx}
\usepackage{txfonts}
\usepackage{lipsum}
\usepackage{subcaption}         
\usepackage{lscape}
\usepackage{placeins}
\usepackage{natbib}
\bibpunct{(}{)}{;}{a}{}{,}

\begin{document}

   \title{Reassessing high-energy emission correlations in gamma-ray bursts using a large, homogeneous sample of X-ray afterglows}

\author{
  A.~A.~Vigliano\inst{1,2}\thanks{\email{alessandro.armando.vigliano@ts.infn.it}}
  \and F.~Longo\inst{1,2}
  \and Ž.~Bošnjak\inst{3}
}

\institute{
  University of Trieste, Department of Physics, via Alfonso Valerio 2, 34127, Trieste, Italy
  \and INFN Trieste, Galleria Padriciano 99, 34149 , Trieste, Italy
  \and University of Zagreb, Faculty of Electrical Engineering and Computing, Unska ul. 3, 10000 Zagreb, Croatia
}

   \date{Received Month Day, 20XX}
 
  \abstract 
   {Gamma-ray bursts (GRBs) exhibit diverse X-ray afterglow light-curves, including breaks and plateau phases, whose physical origins remain debated.
Previous studies have often relied on small and heterogeneous samples, which limited their statistical power and led to apparently conflicting claims.
Most notably, correlations have been suggested between high-energy ($E\ge100$ MeV) detection and X-ray afterglow complexity or plateau incidence.}
   {We performed a comprehensive large-scale and unbiased statistical analysis of GRB X-ray afterglows to clarify the origins of light-curve complexity and plateau phases, and to reassess their possible connection with high-energy emission.}
   {We developed a fully automated, model-independent analysis of the complete \textit{Swift}-XRT GRB afterglow catalog, comprising more than 1400 events.
Our pipeline applies uniform flare removal and segmented power-law fitting, enabling consistent measurements throughout the entire sample and robust statistical inference.}
   {We found that both light-curve complexity and plateau incidence are strongly governed by the data-taking starting time of XRT observations ($t_{XRT}$).
When $t_{XRT}$ is ignored, apparent correlations emerge between high-energy emission and X-ray afterglow morphology, but when the data are stratified or controlled for $t_{XRT}$, these associations vanish.
X-ray complexity and plateaus are therefore not directly coupled to high-energy detectability, indicating that early X-ray morphological features are not predictive of high-energy emission.}
   {Our results resolve long-standing claims in the literature and highlight a crucial methodological lesson: controlling for $t_{XRT}$ is indispensable in large-sample studies of GRBs.
The automated analysis pipeline provides a foundation for reproducible, large-sample inferences and will be essential for exploiting upcoming datasets from missions such as SVOM, Einstein Probe and THESEUS.}


\authorrunning{A.~A.~Vigliano et al.}
\titlerunning{GRB high-energy emission correlations}
   \maketitle
   \nolinenumbers

\section{Introduction}
\label{sec:intro}
Gamma-ray bursts (GRBs) are among the most powerful cosmic explosions.
They produce prompt and afterglow emissions that span the entire electromagnetic spectrum (see \cite{2024Univ...10...57V} for a recent review).
The X-ray afterglow phase, in particular, displays a rich phenomenology: steep early decays, plateau phases, flares, and late-time breaks.
Although a general canonical shape has been proposed \citep{2006ApJ...642..389N, 2007ApJ...662.1093W, 2006ApJ...642..354Z}, individual light-curves often diverge significantly, and the physical origins of several features, such as plateaus and flares, remain a matter of debate.
The standard interpretation of the X-ray afterglow is based on the external shock model, in which the forward shock produced by the relativistic ejecta interacting with the circumburst medium gives rise to the broadband synchrotron emission that dominates at late times.
However, observations from the Neil Gehrels Swift Observatory X-ray Telescope (\textit{Swift}-XRT) have revealed early-time (within about a few hundred to a few thousand seconds after the GRB trigger) features such as steep decays, plateaus, and flares that deviate from the expectations of a simple external-shock scenario \citep{2006ApJ...642..389N, 2006ApJ...647.1213O, 2006ApJ...642..354Z}.
These features require additional physical ingredients, including long-lasting central-engine activity \citep{2013MNRAS.428..729M}, long-lived reverse shocks, or structured jets viewed off-axis.
In particular, plateau phases have been modeled as manifestations of late energy injection or magnetar spin-down processes, while flares are generally attributed to episodic central-engine activity \citep{2006ApJ...647.1213O, 2006ApJ...642..354Z, 2013MNRAS.428..729M}.

The most frequently discussed signatures include plateau phases, also referred to as shallow decay phases, and light-curve complexity.
Both have been suggested as important diagnostics of central engine emissions and jet structure.
These might be used as possible markers of connections to high-energy (HE, $E\ge100$ MeV) and very high-energy (VHE, $E\ge100$ GeV) emission detected by instruments such as the Fermi Gamma-ray Space Telescope Large Area Telescope (\textit{Fermi}-LAT) and ground-based Cherenkov telescopes.
Recently, some studies have reported that GRBs with HE photons tend to show less complex X-ray afterglows or are less likely to display plateaus (e.g. \cite{2020MNRAS.494.5259Y, 2022NatCo..13.5611D}).
If this were confirmed, these correlations would have important implications for the interplay between internal- and external-shock processes in GRBs and for theoretical models of HE and VHE emission (e.g. \cite{2011ApJ...732...77M, 2024ApJ...970..141A}).
A correlation might indicate a common underlying geometric origin for these features, such as a shared dependence on jet-opening angle, viewing angle, or jet structure.
The presence of plateaus (e.g. \cite{2020MNRAS.492.2847B}) and the detection of HE emission (e.g. \cite{2024MNRAS.528.4307B}) have both been proposed in the literature to depend on such geometric parameters.

However, GRB afterglow studies have historically faced two limitations.
First, most analyses were based on relatively small samples (typically 60–80 GRBs; e.g. \cite{2025ApJ...985..261D}), which limits statistical power and comparability.
Second, the lack of uniform analysis procedures (in particular, the fitting of light-curves and the identification of features such as flares and plateaus) makes it difficult to establish whether the reported differences reflect astrophysical trends or methodological choices.
A further challenge lies in the role of the data-taking starting time $t_{XRT}$, which is defined as the elapsed time between the GRB trigger time and the moment \textit{Swift}-XRT started taking data for it.
Because earlier $t_{XRT}$ values naturally encompass a wider temporal baseline, they can artificially increase the measured complexity and alter the detectability of plateaus.
If this is not accounted for, $t_{XRT}$ can act as a hidden confounder in correlations between afterglow features and HE emission.

To overcome these issues, we developed a fully automated, model-independent pipeline for the systematic analysis of X-ray afterglow light-curves.
The pipeline applies a uniform flare removal and segmented linear regression in log–log space, yielding reproducible measurements of breakpoints, plateaus, and other temporal properties for the entire \textit{Swift}-XRT catalog.
Applied to 1438 GRBs observed between December 2004 and December 2025, this represents by far the largest homogeneous study of X-ray afterglows to date.
The scale and uniformity of the analysis enable us for the first time to apply rigorous statistical tests to assess the proposed correlations with HE emission.

We focus on two relations that have been suggested in the literature: (i) between HE emission and X-ray afterglow complexity, and (ii) between HE emission and X-ray plateau phases.
We show that these two apparent associations are in fact driven by the distribution of $t_{XRT}$.
When this is controlled for, neither complexity nor plateau incidence shows a significant link with HE emission.
This result is consistent with scenarios in which the processes shaping early X-ray morphology do not strongly regulate HE emission detectability, and it demonstrates the importance of incorporating $t_{XRT}$ control in all future large-sample studies.

The remainder of this paper is organized as follows.
In Section~\ref{sec:sample}, we describe the dataset, and in Section~\ref{sec:pipeline} we describe the automated analysis pipeline.
Section~\ref{sec:groups} presents the results for complexity and plateaus, with and without $t_{XRT}$ control.
In Section~\ref{sec:physics}, we interpret the findings in the context of GRB emission models, and Section~\ref{sec:conclusions} summarizes our main conclusions and implications for future work.

\section{Sample of GRBs}
\label{sec:sample}
The analysis we present is based on the complete set of X-ray afterglows observed by \textit{Swift}-XRT from December 2004 to December 2025.
Over this period, XRT followed up 1565 GRBs and produced the most extensive and homogeneous archive of X-ray afterglow light-curves to date.
For this study, we initially selected 1438 events with sufficient photon statistics (section \ref{data_prep}) and temporal coverage for a reliable light-curve fitting.
We analyzed flux light-curves between 0.3 and 10 keV (derived using a constant energy conversion factor) obtained from the \textit{Swift}-XRT repository of GRBs\footnote{\url{https://www.swift.ac.uk/xrt_curves/}} \citep{2007A&A...469..379E, 2009MNRAS.397.1177E}.

In order to test the proposed correlations between X-ray light-curve morphologies and the presence of HE emission (Section~\ref{sec:groups}) we limited the sample to events observed after 2008.
From the cleaned sample of 1414 events (Section~\ref{sec:fit}), this resulted in 1215 light-curves.
We then divided them into two groups: a group with and another group without a HE detection.
To define the group of GRBs with a detection at HE we took all the events reported in the second \textit{Fermi}-LAT GRB catalog on HEASARC\footnote{\url{https://heasarc.gsfc.nasa.gov/w3browse/fermi/fermilgrb.html}} from 2008 to 2022 \citep{2019ApJ...878...52A} and added all the GRBs for which a General Coordinates Network (GCN) circular of detection by \textit{Fermi}-LAT has been published up to December 2025.
The seven GRBs seen by the Major Atmospheric Gamma Imaging Cherenkov Telescopes (MAGIC), High Energy Stereoscopic System (H.E.S.S.) and Large High Altitude Air Shower Observatory (LHAASO) observatories in the VHE (GRB 160821B, GRB 180720B, GRB 190114C, GRB 190829A, GRB 201015A, GRB 201216C and GRB 221009A) were also included.
In this way, we obtained a sample of 92 and 1123 GRBs with and without a HE/VHE detection, respectively.

In order to verify that the distribution of $n_{breaks}$ for GRBs without HE detections was not biased by \textit{Fermi}-LAT non-observations, we repeated the analysis in Section~\ref{sec:groups} using a restricted sample of 54 GRBs from \cite{2016ApJ...822...68A}, which were inside the \textit{Fermi}-LAT field of view at trigger time but were not detected by \textit{Fermi}-LAT.

\section{Automated analysis pipeline}
\label{sec:pipeline}
Extracting robust inferences from such a large dataset required a uniform and scalable analysis strategy.
Manual or semi-manual approaches that were typically adopted in earlier studies, are prone to subjective choices, inconsistent fitting procedures, and limited throughput.
This has restricted most previous analyses to relatively small samples (on the order of tens of bursts), making it difficult to establish statistically significant correlations or to compare results across studies.

To overcome these limitations, we developed a fully automated model-independent pipeline for the analysis of \textit{Swift}-XRT afterglows.
The pipeline applies a consistent flare removal followed by segmented linear regression in log–log space.
This allowed us to describe the afterglow as a series of power-law segments.
From this representation, we automatically extracted key features such as the number of temporal breaks (a measure of light-curve complexity) and the presence and duration of plateau phases.
The entire process was designed to be lightweight and reproducible to ensure that identical procedures were applied to every burst in the sample.

The advantages of automation extend beyond the present dataset.
With current and new missions such as the Space Variable Objects Monitor (SVOM), Einstein Probe, and Transient High-Energy Sky and Early Universe Surveyor (THESEUS) expected to greatly increase the number of detected GRBs, reproducible automated pipelines will be essential for the timely and homogeneous characterization of afterglows.
Our approach demonstrates how such tools can transform raw light-curve data into physically meaningful parameters at scale, enabling rigorous statistical tests of proposed correlations between afterglow features and HE emission.

\subsection{Data preparation}
\label{data_prep}
First, the pipeline loads light-curve data without including any upper limit (we did not consider upper limits in this work), and checks that the number of data points is greater than a minimum threshold.
When the number of data points falls below this threshold (set by default to 5), the light-curve is rejected.

Next, the data are converted into log-log space.
This allows the use of linear interpolation techniques instead of a direct power-law fit and greatly reduces computational costs and run times.
It also means that bulk application to large sets of events is possible.

\subsection{Flare identification and removal}
In order to be fully model independent in the flare identification and to avoid introducing bias in the subsequent fit of the afterglow light-curve, a fully data-driven strategy was developed.

\begin{figure}
    \begin{center}
        \includegraphics[width=\hsize]{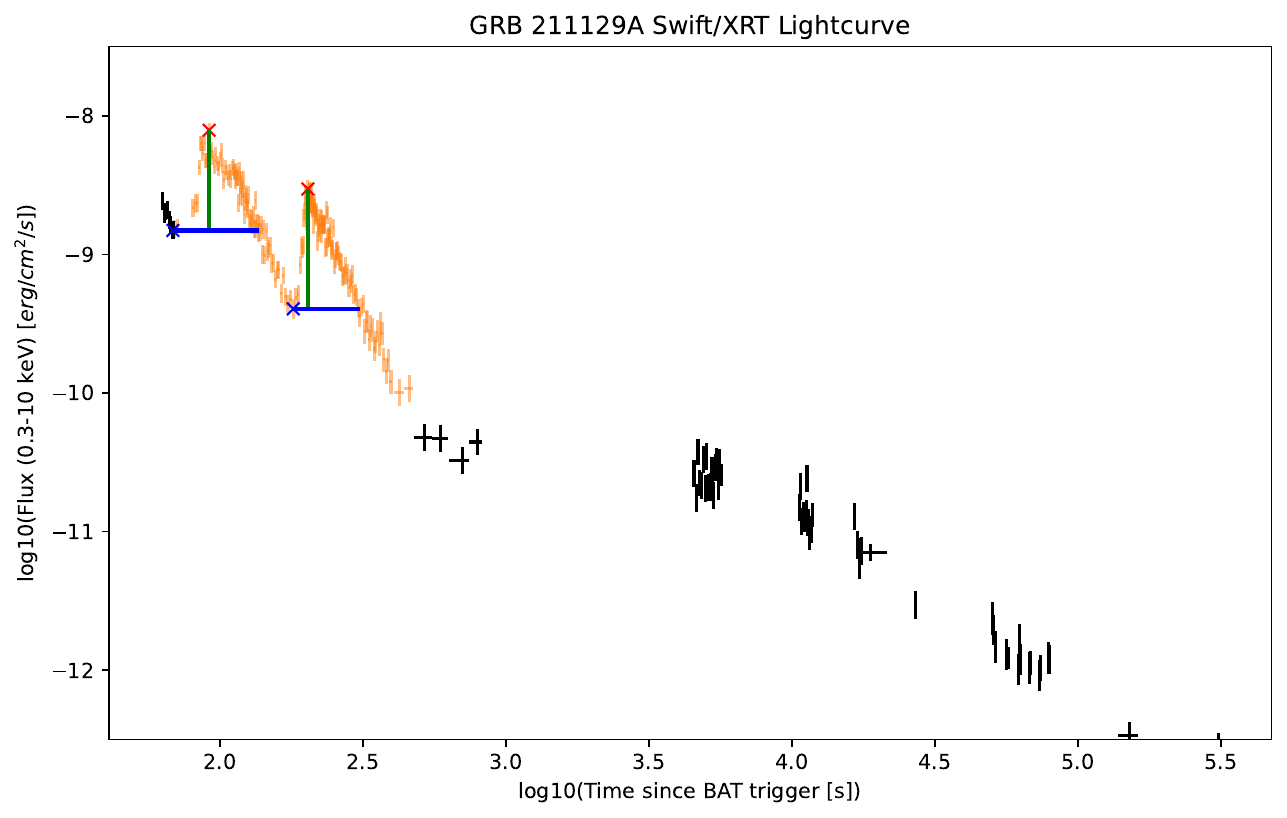}
    \end{center}
    \caption{Each peak with a minimum prominence (vertical green lines) and width (horizontal blue lines) is identified. When the peak significance (the distance between the peak and the base divided by the errors summed in quadrature) exceeds a minimum threshold and the ascending portion contains a minimum fraction of data points, then the peak is identified as a flare. Other considerations are used to limit false negatives for early flares.}
    \label{fig:flareraise}
\end{figure}

To this end, the flare identification pipeline was based on only one assumption: that, except for flaring events, the general trend of the light-curve is decreasing.
This allowed us to identify and perform a first selection of any peaks in the light-curve using their prominence and width.
The prominence of a peak is defined as the height of the peak above the lowest contour line encircling it without containing any higher peaks (blue lines in Fig. \ref{fig:flareraise}).
In this context, these two properties are always well defined, and in particular, the prominence can be identified as the height above the preceding valley closest to the peak (here called the left base of the peak).
This automatically provides an estimate of the onset of flaring activity (Fig. \ref{fig:flareraise}).
To be identified as flares the selected peaks needed to fulfill the following criteria:\begin{enumerate}
\item The significance of the peak from the left base, defined as the difference of the fluxes (in lin-lin space) divided by the errors added in quadrature, must be above a threshold. This was fixed at $3\sigma$ here.
\item The peak must either\begin{itemize}
    \item have a minimum amount of signal (data fraction) in the ascending part greater than a threshold (set in this analysis to 0.7\% of the total number of data points in the light-curve)
    \item occur before this threshold is reached (early flare).
\end{itemize}  
\end{enumerate}
The values of these thresholds were obtained through trials and improvements to minimize the number of false positives and negatives.
These criteria were inspired by those described in \cite{2009MNRAS.397.1177E}.

\begin{figure}
    \begin{center}
        \includegraphics[width=\hsize]{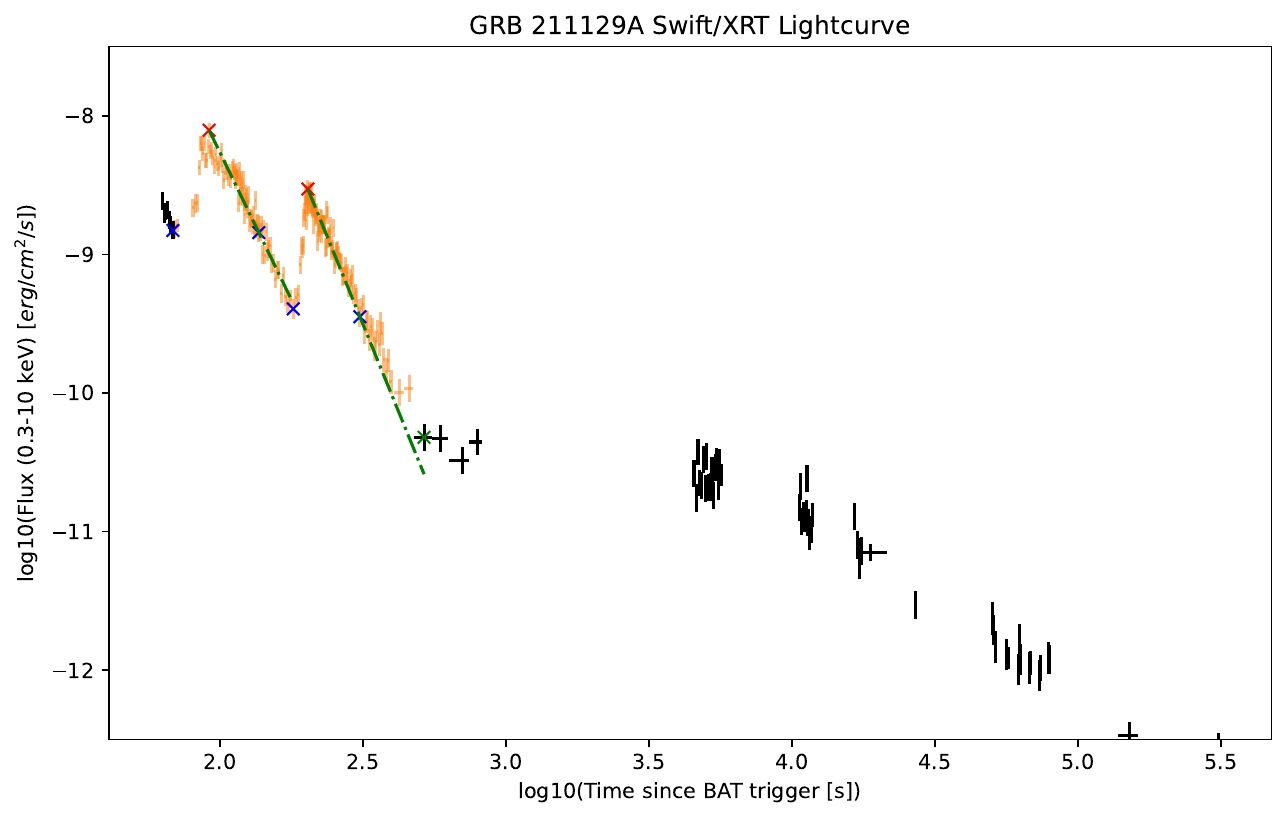}
    \end{center}
    \caption{Tail end of each flare, determined by comparing post-flare points to a reference slope (dash-doted green line) between the peak (red cross) and the first point at the same flux level as the onset of the flare (blue cross). The flare is considered to have ended when the deviations exceed $5\sigma$ (green cross).}
    \label{fig:flaretail}
\end{figure}

To determine the tail end of each flare, two criteria were used:\begin{enumerate}
    \item A flare was considered to have ended when the time lag between successive observations was larger than a threshold in log-log space.
    \item Post-peak points were compared to a reference power-law slope passing through the peak and the first point at the same flux level of the beginning of the flare (the left base of the flare).
    The flare was considered to have ended when the deviations, smoothed using a moving-window (rolling median) approach, exceeded $5\sigma$ (Fig. \ref{fig:flaretail}):\begin{equation}
        \text{Flare\_End}(y)=\{i\,|r_i\geq5\sigma\},\,r_i=\text{median}\Bigl(\frac{y_i-\hat{y}_i}{\sigma_{y_i}}\Bigr)_{i-2}^{i+2}
    \end{equation}
    where $\hat{y}_i$ is the reference power-law value at time i.
    A rolling median makes the tail identification more robust by eliminating the risk of cutting the flare short due to large random oscillations caused by noise.
\end{enumerate}

\begin{figure}
    \begin{center}
        \includegraphics[width=\hsize]{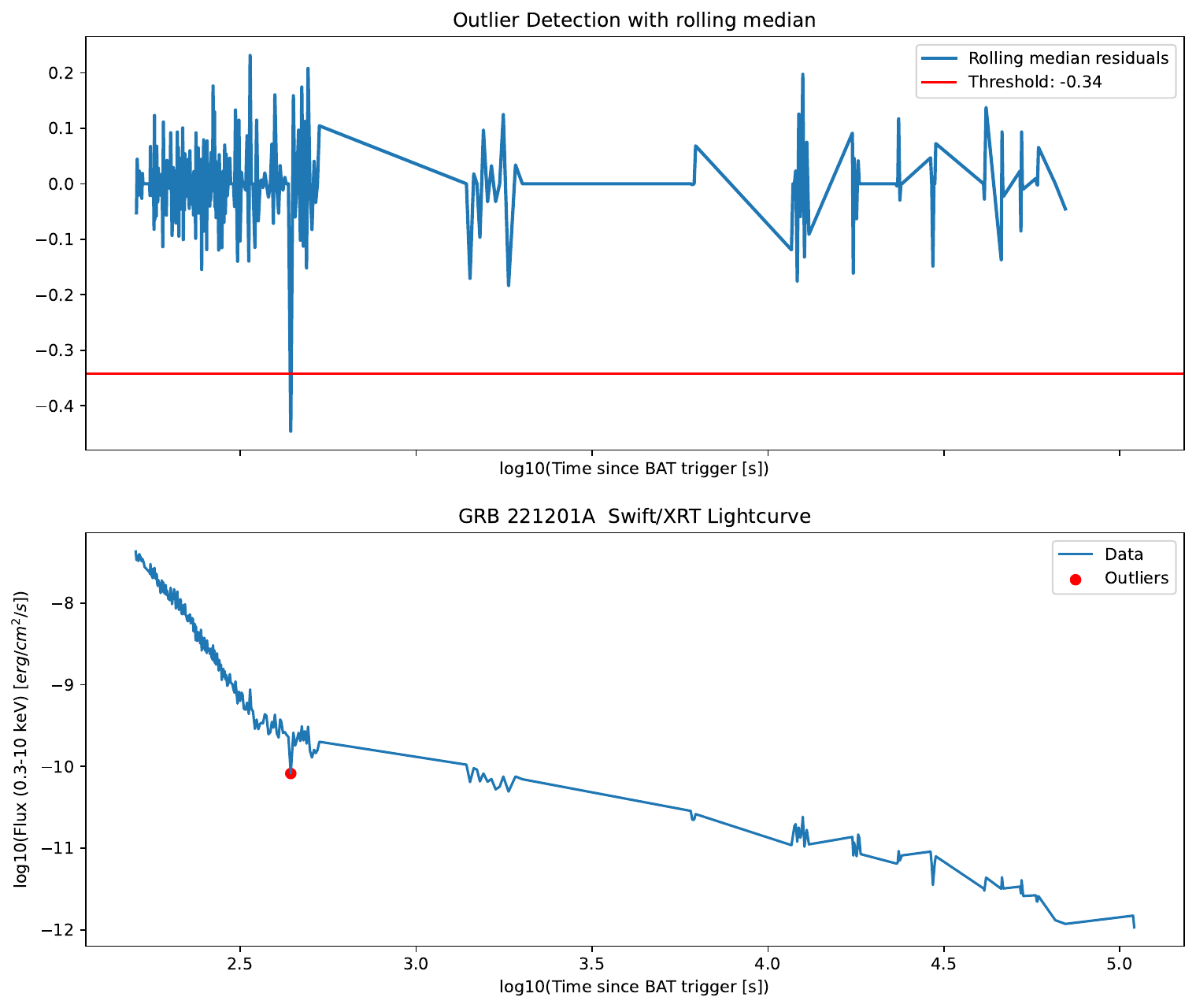}
    \end{center}
    \caption{Outliers were identified with a fixed moving-window approach (three consecutive data points in index space), with which we evaluated the difference between each data point and the local rolling median (top panel). Any points where the downward deviation was $\leq-5\sigma$ were flagged as significant downward outliers (red marker, bottom panel). These points were excluded from the flare identification procedure to prevent significant random downward fluctuations from being misidentified as flare onsets.}
    \label{fig:outlierflares}
\end{figure}

Finally, to limit the number of false positives, a list of possible negative outliers was obtained.
To do this, a simple moving-window approach was used to evaluate the local behavior of the data.
For each data point, the difference between its value and the moving median calculated over a window of three points centered on that point was evaluated.
Any points for which the deviation was smaller than or equal to the negative of a threshold (here set at $5 \sigma$) were considered significant downward outliers (Fig. \ref{fig:outlierflares}):\begin{multline}
\text{Outliers}(y)=\{i\,|\,r_i\leq -5\sigma_r\},\,r_i=y_{i}-\text{median}(y_{i-1},y_{i},y_{i+1})
\end{multline}
This procedure does not assume a contiguous exposure and remains robust to irregular sampling.
The moving-window was defined in index space and operated on consecutive data points, regardless of their temporal separation.
\textit{Swift}-XRT light-curves contain orbital gaps and irregular sampling, but the algorithm compares each point only with its immediate neighbors in the series.
Under the assumption that, outside flaring episodes, the afterglow exhibits a generally monotonic decay, the method flags only statistically significant downward deviations relative to the local median of adjacent measurements.
Monotonic decay, even across temporal gaps, therefore does not produce spurious detections.
Instead, the procedure was designed specifically to identify isolated significant random downward fluctuations that could otherwise be misidentified as flare onsets.

This analysis identified at least one flare in 426 GRBs ($\sim29.61\%$ of the sample) we analyzed from the \textit{Swift}-XRT GRB catalog \citep{2007A&A...469..379E, 2009MNRAS.397.1177E}, in line with the values generally found in the literature ($\approx30\%$, \cite{2019ApJ...884...59L}).
\begin{figure*}
    \centering
    {\includegraphics[width=\textwidth]{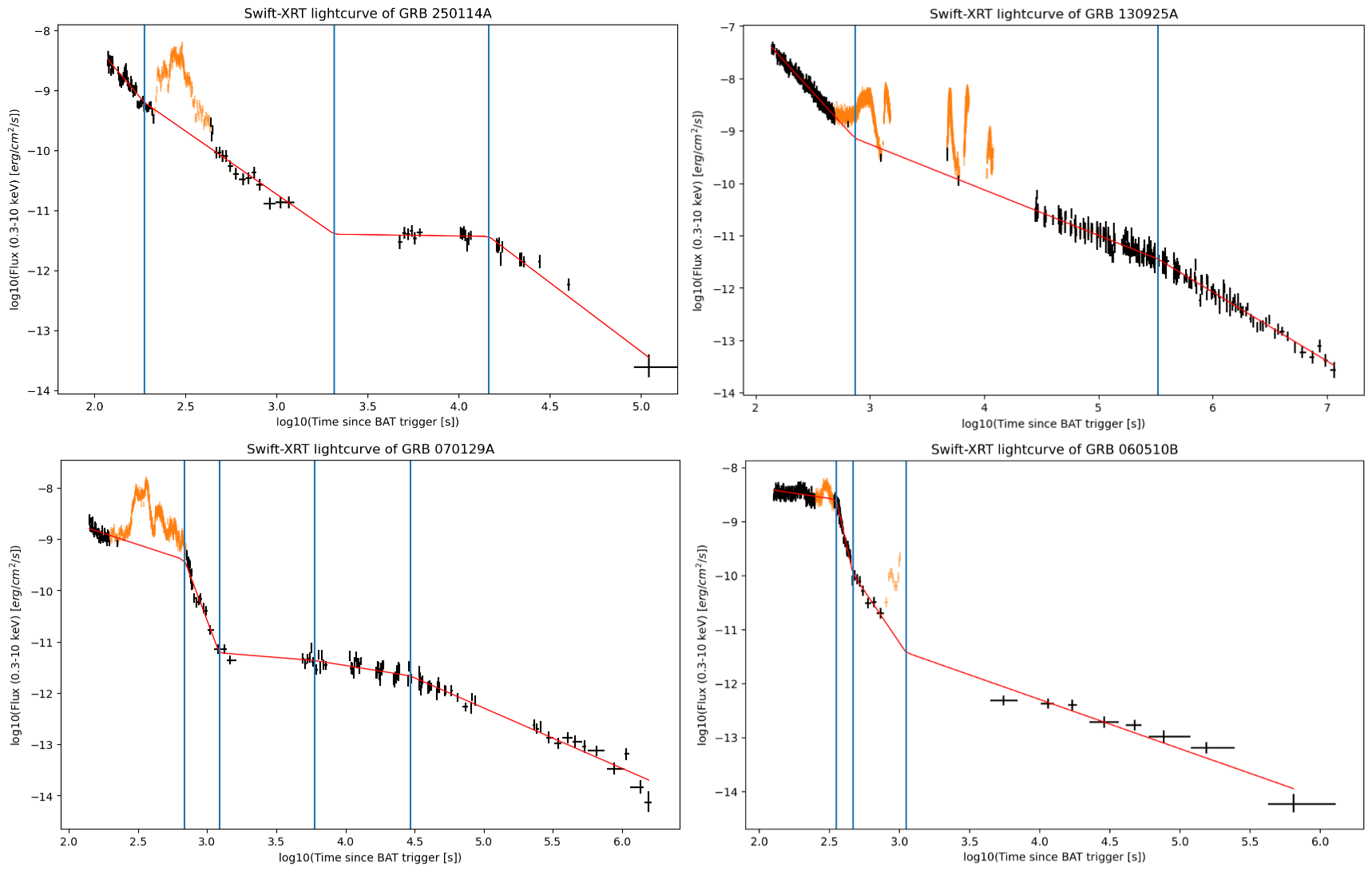}}
     \caption{Some examples of GRB light-curves analyzed with this new automatic analysis. The flares are identified and removed (orange). The breakpoints are reported as vertical blue lines.}
      \label{fig:grbesxamples}
\end{figure*}
\subsection{Afterglow fit and model selection}
\label{sec:fit}
After flares were removed from the light-curves, the light-curves were fitted with various functions to characterize the general behavior of the afterglow.
For each bin, we used the time stamp provided in the \textit{Swift}-XRT light-curve files (i.e., the representative time assigned to that bin).
Since each point represents an average over a finite time interval, a fully rigorous treatment would integrate the model over each bin.
However, this refinement is not expected to affect our primary classification metrics (break multiplicity and plateau incidence), and we therefore adopted the standard approach of fitting against the provided bin time stamps.
The time sampling of the Swift-XRT light-curves is determined by the adaptive binning adopted in the Leicester pipeline, which optimizes the signal-to-noise ratio of each measurement. 
In principle a finer temporal binning could be produced, but this would significantly increase the statistical uncertainties, especially at late times when the count rate is low.

In order to compare the results with the official Swift automatic analysis \citep{2009MNRAS.397.1177E}, the light-curves were fitted with different power-law segments with up to five breakpoints.
This also agrees with the main expectations from the GRB X-ray canonical light-curve (\cite{2006ApJ...642..354Z, 2007ApJ...662.1093W}).
However, to limit the risk of overfitting in the case of very few data points, the maximum number of breakpoints was limited so that the number of data points was at least twice (editable parameter) the number of free parameters of the model.

The use of data in log-log space reduces the multiple power-law fit to a segmented regression problem using straight lines (Fig. \ref{fig:grbesxamples}).
An excellent and well-tested method to do this was proposed by \cite{Muggeo03}.
This approach fits the data with a piecewise linear function, allowing for unknown breakpoints where the slope may change.
For a single independent variable and N breakpoints, the model takes the form:\begin{equation}
    y=c+\alpha_1x+\sum_{k=1}^N\beta_{k}(x-\psi_k)H(x-\psi_k)+\epsilon,
\end{equation} where $\alpha_1$ is the slope of the first segment, $c$ is the intercept of the first segment, $H$ is the Heaviside step function, $\psi_k$ are the estimated breakpoint locations,  $\beta_k$ represent changes in slope at each breakpoint and $\epsilon$ is a noise term.
This method iteratively estimates the regression coefficients and the breakpoints via a linearization-based updating scheme.
A Taylor series expansion is performed around an estimate of each breakpoint, making the dependence on the actual breakpoint position linear.
This allow us to evaluate a new, improved estimate of the breakpoint position with a simple linear regression, which can then be used to repeat this analysis until the estimate converges to the actual position.
This algorithm was implemented using a custom version of the \texttt{Piecewise-regression} Python library \citep{Pilgrim2021}.

Finally, in order to select the best-fitting model, different statistics were implemented.
The \texttt{Piecewise-regression} library comes with a tool for comparing the models by minimizing the Bayesian information criterion (BIC).

In a few cases (24 events), the combination of limited data points and segmented fitting produced unphysically extreme slopes.
This is indicative of residual overfitting despite the constraints described above.
These light-curves were excluded from the final sample to avoid bias from poorly constrained fits, leaving a total of 1414 events that were used in the statistical analysis prior to any subsample selections.
This corresponds to less than $2\%$ of the dataset and does not affect the statistical results presented in this work.

\section{Testing GRB HE emission correlations}
\label{sec:groups}
With the automated pipeline applied to the entire \textit{Swift}–XRT catalog, we obtained homogeneous measurements of light-curve complexity (number of breaks) and plateau (defined as power-law segments with $-0.7\le\alpha\le0.3$) incidence for 1414 GRBs.
In order to compare these features between bursts with and without detected high-energy (HE) emission we limited the sample to the GRBs that were observed after the launch of the Fermi mission (2008), as described in Section~\ref{sec:sample}.

These two properties have been proposed in the literature as key signatures of the connection between jet structure and HE emission in limited samples of GRBs ($\lesssim$ 30 events).
In what follows, we first examine the raw group differences, and we then test whether they persist when $t_{XRT}$ is taken into account.
\subsection{Complexity versus high-energy detection}
\begin{figure*}
    \centering
    {\includegraphics[width=\textwidth]{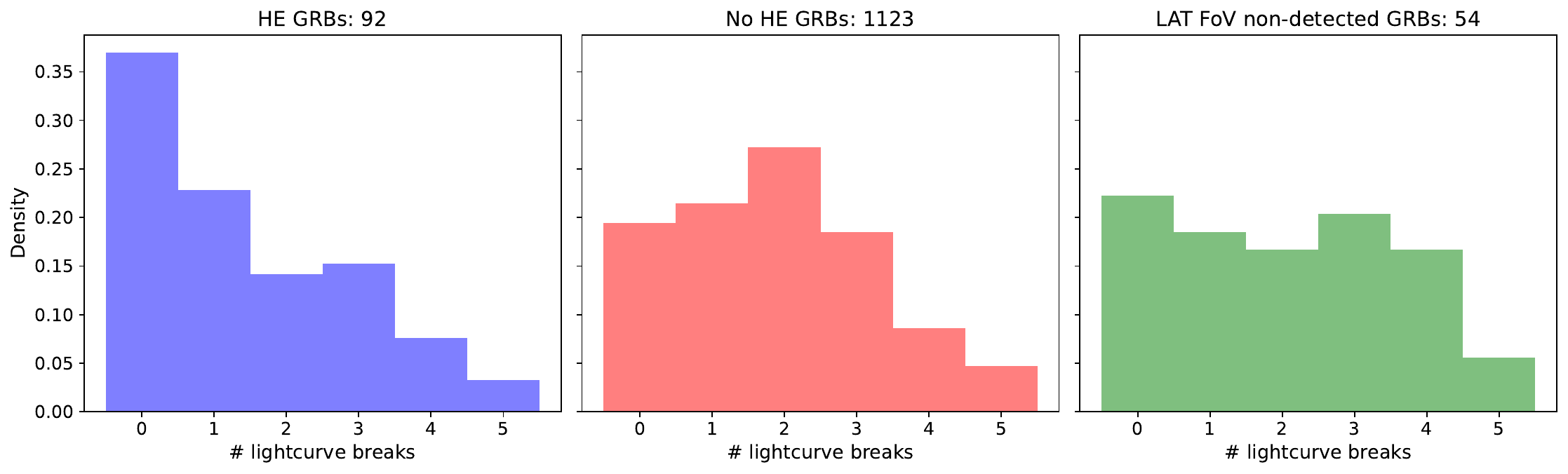}}
     \caption{Comparison of the distributions of the number of breakpoints for three groups of GRBs: GRBs with a detection at HE/VHE (blue, left), GRBs observed in X-rays, without a detection at HE (red, center), and GRBs observed in X-rays, without a detection at HE, but inside the \textit{Fermi}-LAT FoV at the moment of triggering (green, right) \citep{2016ApJ...822...68A}. This group was used as a cross-check to verify that the distribution of $n_{breaks}$ for GRBs without HE detections (in red) was not biased by \textit{Fermi}-LAT non-observations.}
      \label{Ackerman}
\end{figure*}
\begin{figure*}
    \centering
    {\includegraphics[width=\textwidth]{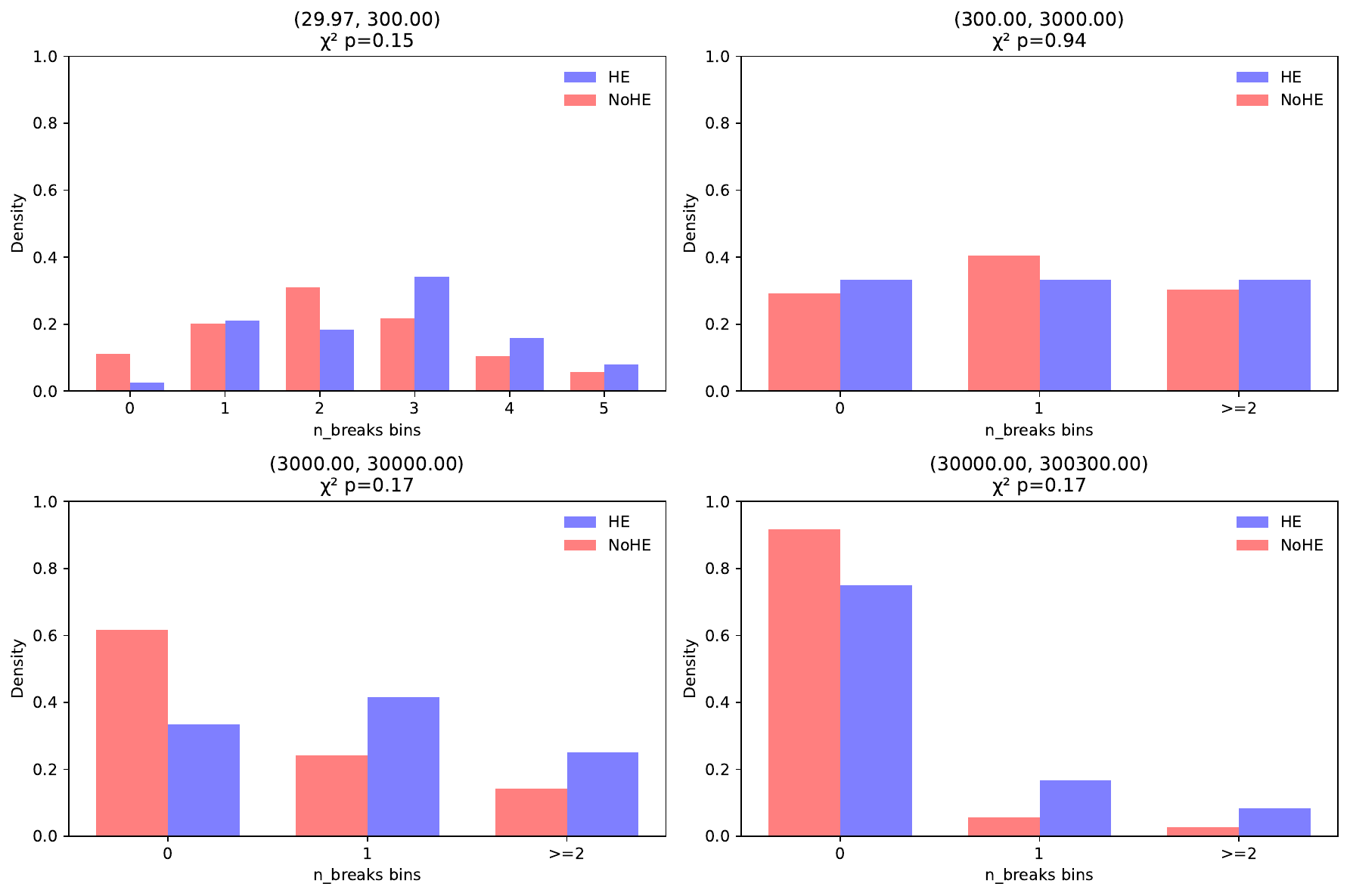}}
     \caption{Comparison of the distributions of the number of breakpoints between GRBs with (blue) and without (red) a HE detection divided in four $t_{XRT}$ bins. Further details are reported in Table \ref{tab:complexity_bins}.}
      \label{Nb_vs_HE_t0}
\end{figure*}
We first compared the distribution of X-ray light-curve complexity, quantified by the number of temporal breaks ($n_{breaks}$) required to fit each light-curve, between GRBs with and without HE detections.
In the raw comparison, HE bursts appeared to be less complex (i.e. they required a lower number of breaks for the best fit), which is consistent with earlier claims in the literature ($\chi^2=19.27, p=1.71\times10^{-3}$, Fig. \ref{Ackerman}).
As a cross-check to verify that the distribution of $n_{breaks}$ for GRBs without HE detections was not biased by \textit{Fermi}-LAT non-observations, we repeated the comparison using a restricted sample of 54 GRBs from \cite{2016ApJ...822...68A} (Section~\ref{sec:sample}) that were inside the \textit{Fermi}-LAT field of view at trigger time but were not detected by \textit{Fermi}-LAT (green plot on the right in Fig. \ref{Ackerman}).
The $n_{breaks}$ distribution for this selected subset is statistically indistinguishable from the full non-HE sample ($\chi^2=6.43, p=0.27$), indicating that the apparent difference in complexity between HE and non-HE GRBs is not driven by observational bias due to incomplete \textit{Fermi}-LAT sky coverage.

To investigate whether this apparent trend depended on the data-taking starting time $t_{XRT}$, we stratified the sample into four logarithmic bins of $t_{XRT}$ and compared the distributions of $n_{breaks}$ in the two groups within each bin (Fig. \ref{Nb_vs_HE_t0}).
As shown in Table \ref{tab:complexity_bins}, stratified analyses across $t_{XRT}$ bins did not reveal statistically significant differences between the two groups.
Chi-square tests yielded p values between 0.15 and 0.94.
Although the Mann–Whitney tests \citep{497e1044-d5b0-30a9-b230-3ca0f10d6f6c} in the earliest interval (30–300 s) produced a nominal p value of 0.033, this did not remain significant after Bonferroni correction for the family of $m=8$ bin-wise tests (four $t_{XRT}$ bins times two statistical tests; adjusted $p_{Bonf}=0.264$, \cite{Bland170}).
The Bonferroni correction was applied across the eight statistical comparisons performed in the stratified analysis, corresponding to two test statistics (chi-square and Mann–Whitney) evaluated in each of the four $t_{XRT}$ bins.
No bin showed a statistically significant difference after adjustment.

A more thorough test was obtained by fitting a Poisson generalized linear model (GLM, \cite{mccullagh2019generalized}) with $n_{breaks}$ as the dependent variable, HE detection as a factor ($\mathrm{HE}=1$ when the GRB had a high-energy detection and 0 otherwise), $\log_{10}(t_{XRT})$ as a covariate and an interaction term:
\begin{multline}
\ln \big( \mathbb{E}[n_{\mathrm{breaks}}] \big) 
= \beta_0 + \beta_{\mathrm{HE}} \cdot \mathrm{HE} 
+ \beta_{t_{XRT}} \cdot \log_{10}(t_{XRT}) \\
+ \beta_{\mathrm{int}} \cdot \mathrm{HE} \times \log_{10}(t_{XRT}),
\end{multline}
where $\mathbb{E}[n_{\mathrm{breaks}}]$ denotes the expected number of breaks.
The model showed that $\log_{10}(t_{XRT})$ strongly predicts complexity ($\beta_{t_{XRT}}=-0.77\pm0.05, p<10^{-15}$): bursts with earlier $t_{XRT}$ values require more breaks for the best fit.
The HE detection factor ($\beta_{HE}=-0.16\pm0.25, p=0.50$) and its interaction with $t_{XRT}$ ($p=0.21$) were not significant.
The model explained about 26\% of the variance (pseudo-R$^2 = 0.28$), with no sign of
overdispersion (Pearson $\chi^2$/d.o.f = 0.87).
The results are reported in Table \ref{tab:GLM_complexity}.

Thus, the apparent X-ray light-curve complexity–HE emission association is consistent with being caused solely by the distribution of data-taking starting times and is not due to intrinsic emission effects of the sources.
The previously suggested link between HE emission and X-ray light-curve complexity is spurious.
When $t_{XRT}$ is taken into account, HE detection provides no additional predictive power.

\subsection{Plateaus versus high-energy detection}
\begin{figure}
    \begin{center}
        \includegraphics[width=\hsize]{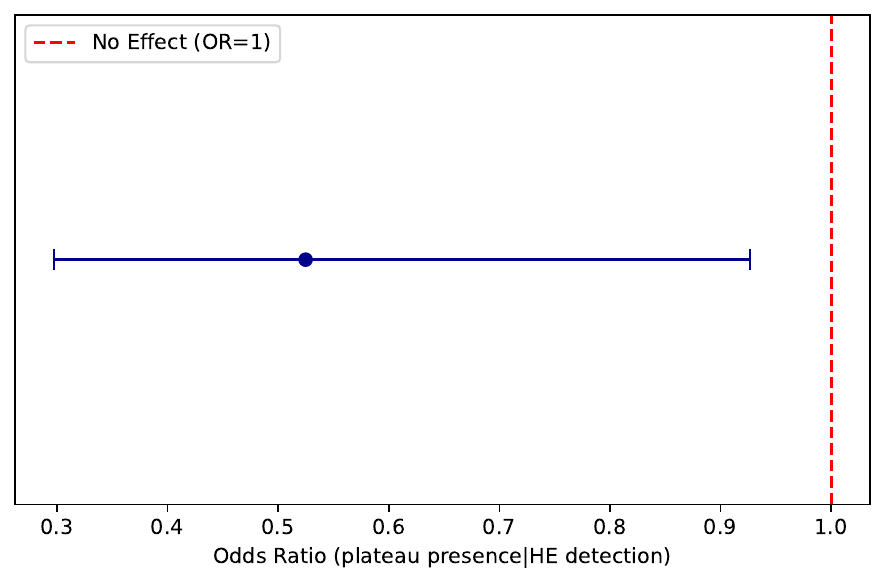}
    \end{center}
    \caption{Forest plot showing the OR and 95\% confidence interval for the association between HE detection and plateau phases in GRBs (OR of plateaus given a HE detection). The vertical dashed red line marks the null hypothesis of no association $(OR = 1)$.}
    \label{platvsHEnaive}
\end{figure}
\begin{figure*}
    \centering
    {\includegraphics[width=\textwidth]{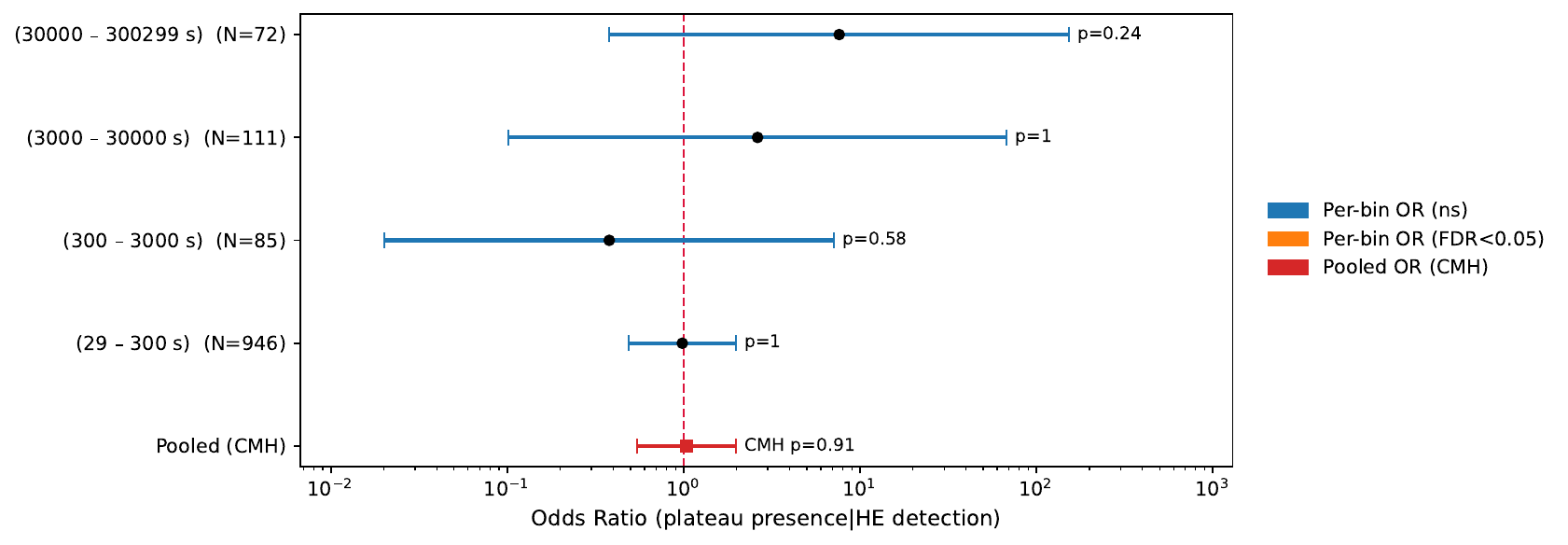}}
     \caption{Forest plot for the association between HE detection and plateau phases in GRBs stratified in bins of $t_{XRT}$ and pooled together with the CMH method. When the data are stratified by $t_{XRT}$ any apparent suppression of plateaus in the HE GRB group disappears with no association between the two variables.}
      \label{platvsHEt0}
\end{figure*}
We next tested whether HE bursts differed in their probability of showing plateau phases.
A raw Fisher exact test on the overall full sample $2\times2$ contingency table suggested a weak association, with HE bursts appearing less likely to display X-ray plateaus (Fig. \ref{platvsHEnaive}).
This would be consistent with earlier claims that GRBs detected at high energies have simpler afterglows with fewer plateau phases.

To account for possible confounding by the data-taking starting time $t_{XRT}$, we repeated the analysis by stratifying the data into four logarithmic bins of $t_{XRT}$ as before (Fig. \ref{platvsHEt0}).  
As summarized in Table~\ref{tab:plateau_bins}, none of the bins shows a significant difference between HE and non-HE samples when $t_{XRT}$ is controlled for.

Combining the four strata using the Cochran–Mantel–Haenszel (CMH) method \citep{54475ba4-6108-364d-b047-e5384517b947, 10.1093/jnci/22.4.719} yielded a pooled-odds ratio consistent with unity ($\mathrm{OR}=1.04$, 95\% CI: 0.54–1.98, $p=0.91$), confirming the absence of a systematic association between plateau incidence and HE detection when $t_{XRT}$ is accounted for (Fig. \ref{platvsHEt0}). 

As a complementary approach we modeled the plateau incidence with a binomial GLM with logit link \citep{Agresti2002}: \begin{equation}
\ln \left( \frac{P(\mathrm{plateau}=1)}{P(\mathrm{plateau}=0)} \right)
= \alpha_0 + \alpha_{\mathrm{HE}} \cdot \mathrm{HE} 
+ \alpha_{t_{XRT}} \cdot \log_{10}(t_{XRT}).
\end{equation}
After we adjusted for $t_{XRT}$, the logistic regression showed no statistically significant difference between GRBs with and without HE detections in the probability of having a plateau phase in this case as well: adjusted OR = 1.00, 95\% CI 0.52-1.92, and p = 0.99 (results in Table \ref{tab:GLM_plat}).

Thus, the apparent anticorrelation in the raw comparison is entirely compatible with being caused by the differences in the $t_{XRT}$ distribution.
After the data-taking starting time is accounted for, the presence or absence of plateau phases in the X-ray light-curves is not associated with HE emission.

\subsection{The impact of $t_{XRT}$}
\begin{figure*}
    \centering
    {\includegraphics[width=\textwidth]{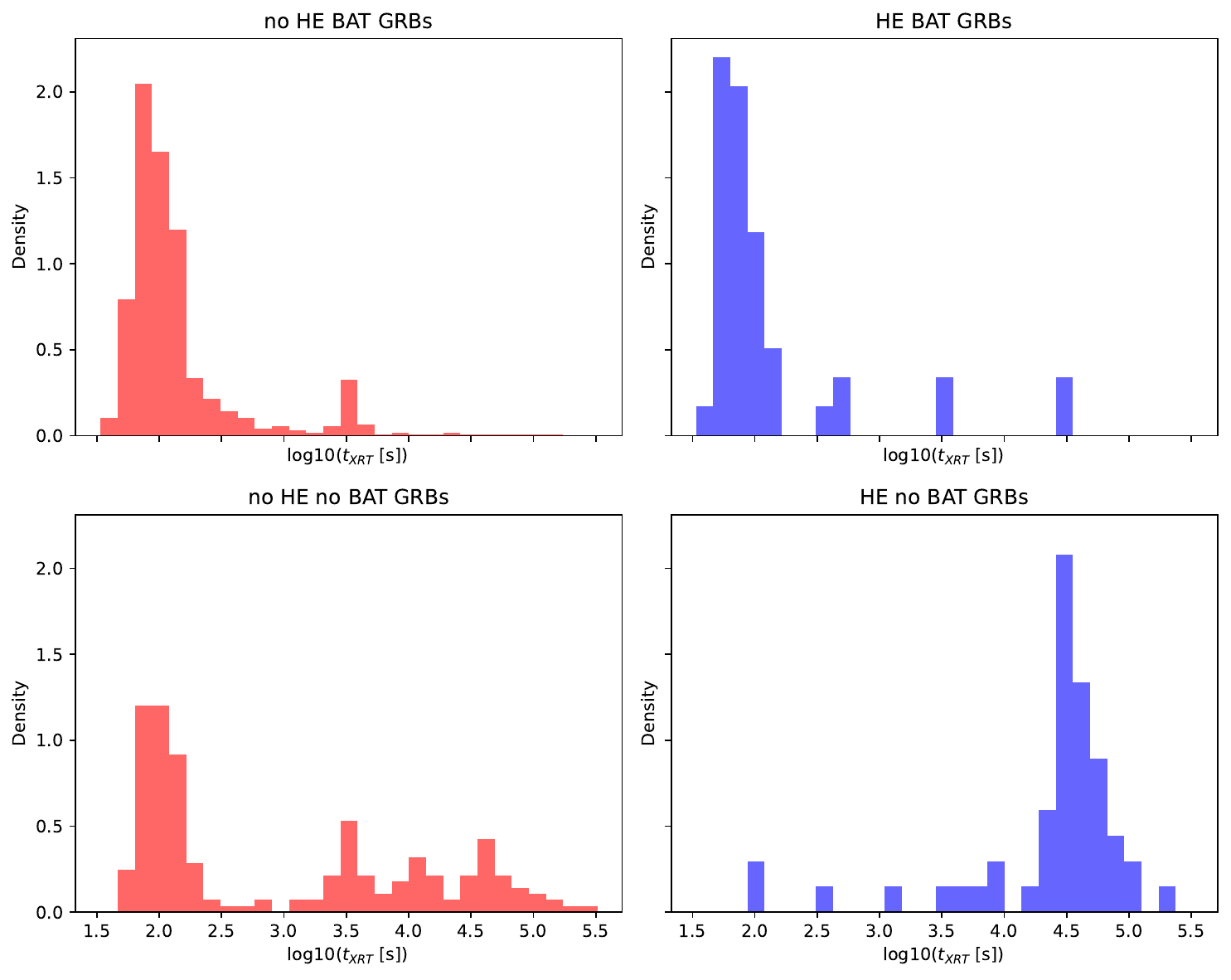}}
     \caption{Histograms of the distribution of $\log_{10}(t_{XRT})$ for the GRB samples with (blue, right) and without (red, left) HE detection, with and without \textit{Swift}-BAT triggering (upper and lower row respectively).}
      \label{fig:t0}
\end{figure*}
The complexity and plateau analyses converge on the same conclusion: neglecting $t_{XRT}$ produces apparent correlations between X-ray afterglow features and HE emission that vanish when $t_{XRT}$ is taken into account.
The data-taking starting time $t_{XRT}$, defined as the elapsed time between the GRB trigger time and the moment \textit{Swift}-XRT started taking data for it, therefore acts as a confounding variable.
Later $t_{XRT}$ values artificially reduce the measured number of breakpoints and the likelihood of identifying a plateau, so that comparing samples with different $t_{XRT}$ distributions inevitably introduces spurious associations.
In particular, the distribution of $t_{XRT}$ for GRBs with HE detection appears to be bimodal with distinct peaks at early ($\sim10^{2}s$) and late ($\sim10^{4.5}s$) timescales (blue column on the right of Fig. \ref{fig:t0}).
This is naturally explained by the operational sequence of Swift and Fermi observations.
GRBs that trigger \textit{Swift}-BAT and are simultaneously within the \textit{Fermi}-LAT field of view typically receive prompt \textit{Swift}-XRT follow-up within minutes.
In contrast, most of the Fermi-LAT GRB detections are identified only after data downlink and validation.
In these cases, a Fermi-LAT offline position notice can trigger a Swift target-of-opportunity (ToO) observation, resulting in Swift-XRT follow-up beginning several hours after the initial trigger.
The resulting mixture of prompt slews and delayed ToO follow-up naturally produces a two-peaked $t_{XRT}$ distribution and reflects an operational selection effect and not an intrinsic GRB timescale.

Crucially, this effect only becomes visible through the scale of our dataset.
By applying a fully automated procedure to the entire \textit{Swift}-XRT catalog, we were able to carry out a genuine very large sample analysis and employ formal statistical inference rather than relying on small, heterogeneous samples.
This shift in method enabled us to disentangle observational biases (such as $t_{XRT}$) from genuine physical trends.

The key methodological lesson is therefore clear: $t_{XRT}$ must be included in any statistical analysis of X-ray afterglow features.
Failure to do so risks drawing misleading conclusions about the physics of GRBs, such as attributing X-ray afterglow complexity or plateau phases to the presence or absence of HE emission when in fact these phenomena are unrelated.

\section{Discussion}
\label{sec:physics}
Our results directly address long-standing claims in the literature regarding a possible connection between high-energy (HE) emission and the morphology of GRB X-ray afterglows.
Previous studies suggested that GRBs with HE detections tend to exhibit less complex early X-ray afterglow light-curves and fewer plateau phases.
With the benefit of a homogeneous, catalog-wide analysis and proper statistical stratification, we found that it is crucial to account for the starting time of \textit{Swift}-XRT observations ($t_{XRT}$) when conclusions about the behavior of HE and non-HE samples are drawn. 
The apparent correlations are entirely driven by differences in $t_{XRT}$ distributions between HE and non-HE samples.

This finding disfavors a simple one-to-one link between early X-ray morphological features (plateau incidence and light-curve complexity) and HE detectability, indicating that the presence of these X-ray features is not predictive of emission at $E\geq100$ MeV when observational biases related to $t_{XRT}$ are taken into account.
The complexity of X-ray afterglows and the presence of plateaus are commonly explained as signatures of central engine activity or other internal processes operating during the early afterglow.
In contrast, the production of HE photons has been linked to external shock physics at later times.
If these observables were physically connected, correlations would emerge in large, uniformly analyzed samples.
After correcting for observational biases (the distribution of $t_{XRT}$) we found no statistically significant association between plateau incidence or light-curve complexity in the X-rays and high-energy detections.
This result indicates that early-time X-ray morphological features are not predictive of emission at $E\geq100$ MeV in current data.
The lack of correlation does not contradict broadband synchrotron models in which X-ray and HE emission can arise from a common external-shock component.
It shows instead that the presence of plateaus or additional breaks in the X-ray light-curves is not directly coupled to the presence of HE emission.
More subtle connections cannot be excluded.
Crucially, the absence of any detectable link in a dataset of this scale resolves earlier conflicting claims and places robust constraints on models attempting to unify early X-ray features with HE/VHE emission.
Earlier and more continuous overlap between HE/VHE and X-ray observations, enabled by future missions and faster response capabilities, will be essential to test whether any physical connection emerges under better-constrained early-time coverage.

The apparent connection reported in previous studies is therefore consistent with arising from observational selection driven by $t_{XRT}$and does not require an intrinsic dependence on HE emission.

Beyond the physical interpretation, our study carries an important methodological message.
Accounting for $t_{XRT}$ is a necessary step in any very large sample analysis of GRB afterglows.
Later start times systematically bias measurements of X-ray light-curve complexity and plateau incidence, and comparisons that ignore this effect inevitably risk producing spurious trends.
Our analysis demonstrates how robust, large-sample inference can overturn earlier claims and clarify the actual drivers of afterglow diversity.
Future GRB studies must therefore explicitly control for $t_{XRT}$ as a standard element of their method.

Additional checks on the properties of the X-ray afterglows, including the distribution of the initial decay slope $\alpha_1$, the comparison between long and short GRBs, and the dependence on redshift, are reported in Appendices~\ref{sec:a1} and ~\ref{sec:redshift}.
The results of these tests are consistent with those presented in the main analysis and do not modify the conclusions.

\section{Conclusions}
\label{sec:conclusions}
We have presented a new fully automated, model-independent analysis of the complete \textit{Swift}-XRT GRB afterglow catalog, yielding homogeneous measurements of light-curve features for more than 1400 bursts.
This approach enabled us for the first time to conduct statistically rigorous large-sample tests of proposed correlations between X-ray afterglow morphology and $E\geq100$ MeV (HE/VHE, as detected by \textit{Fermi}-LAT and ground-based Cherenkov telescopes) emission.
Our main conclusions are listed below.\begin{enumerate}
    \item The previously reported association between HE emission and X-ray afterglow light-curve complexity is an artifact of not considering the observation starting time $t_{XRT}$.
    When they are stratified by or controlled for $t_{XRT}$, HE bursts are no less complex than no-HE bursts.
    \item Similarly, the apparent deficit of X-ray plateau phases in HE bursts disappears when $t_{XRT}$ is accounted for.
    \item Together, these results indicate that plateau incidence and light-curve complexity are not statistically associated with HE emission detection. While broadband models can accommodate common emission origins, our findings showed that X-ray morphological features are not directly linked to HE detectability.
\end{enumerate}
Methodologically, the work highlights the power of automated analysis pipelines for enabling reproducible and statistically sound large-sample studies.
It also established a key lesson for the field: control for $t_{XRT}$ is indispensable.
Ignoring it leads to biased comparisons and incorrect physical conclusions.
Looking ahead, the automated framework developed here provides a blueprint for analyzing the large datasets expected from current and future missions such as SVOM, Einstein Probe and THESEUS.
With proper treatment of $t_{XRT}$, these missions will allow unprecedented insight into the multiwavelength physics of GRB afterglows.

\begin{acknowledgements}
This work made use of data supplied by the UK Swift Science Data Centre at the University of Leicester.
\end{acknowledgements}

\bibliographystyle{aa}
\bibliography{references.bib}

@ARTICLE{2024Univ...10...57V,
       author = {{Vigliano}, Alessandro Armando and {Longo}, Francesco},
        title = "{Gamma-ray Bursts: 50 Years and Counting!}",
      journal = {Universe},
         year = 2024,
        month = jan,
       volume = {10},
       number = {2},
          eid = {57},
        pages = {57},
          doi = {10.3390/universe10020057},
       adsurl = {https://ui.adsabs.harvard.edu/abs/2024Univ...10...57V}
}

@article{Muggeo03,
author = {Muggeo, Vito M. R.},
title = {Estimating regression models with unknown break-points},
journal = {Stat. Med.},
volume = {22},
number = {19},
pages = {3055-3071},
doi = {https://doi.org/10.1002/sim.1545},
url = {https://onlinelibrary.wiley.com/doi/abs/10.1002/sim.1545},
eprint = {https://onlinelibrary.wiley.com/doi/pdf/10.1002/sim.1545},
year = {2003}
}

@article{Pilgrim2021, doi = {10.21105/joss.03859}, url = {https://doi.org/10.21105/joss.03859}, year = {2021}, publisher = {The Open Journal}, volume = {6}, number = {68}, pages = {3859}, author = {Charlie Pilgrim}, title = {piecewise-regression (aka segmented regression) in Python}, journal = {JOSS} }

@ARTICLE{2009MNRAS.397.1177E,
       author = {{Evans}, P.~A. and {Beardmore}, A.~P. and {Page}, K.~L. and {Osborne}, J.~P. and {O'Brien}, P.~T. and {Willingale}, R. and {Starling}, R.~L.~C. and {Burrows}, D.~N. and {Godet}, O. and {Vetere}, L. and {Racusin}, J. and {Goad}, M.~R. and {Wiersema}, K. and {Angelini}, L. and {Capalbi}, M. and {Chincarini}, G. and {Gehrels}, N. and {Kennea}, J.~A. and {Margutti}, R. and {Morris}, D.~C. and {Mountford}, C.~J. and {Pagani}, C. and {Perri}, M. and {Romano}, P. and {Tanvir}, N.},
        title = "{Methods and results of an automatic analysis of a complete sample of Swift-XRT observations of GRBs}",
      journal = {\mnras},
         year = 2009,
        month = aug,
       volume = {397},
       number = {3},
        pages = {1177-1201},
          doi = {10.1111/j.1365-2966.2009.14913.x},
archivePrefix = {arXiv},
       eprint = {0812.3662},
 primaryClass = {astro-ph},
       adsurl = {https://ui.adsabs.harvard.edu/abs/2009MNRAS.397.1177E}
}

@ARTICLE{2007ApJ...662.1093W,
       author = {{Willingale}, R. and {O'Brien}, P.~T. and {Osborne}, J.~P. and {Godet}, O. and {Page}, K.~L. and {Goad}, M.~R. and {Burrows}, D.~N. and {Zhang}, B. and {Rol}, E. and {Gehrels}, N. and {Chincarini}, G.},
        title = "{Testing the Standard Fireball Model of Gamma-Ray Bursts Using Late X-Ray Afterglows Measured by Swift}",
      journal = {\apj},
         year = 2007,
        month = jun,
       volume = {662},
       number = {2},
        pages = {1093-1110},
          doi = {10.1086/517989},
archivePrefix = {arXiv},
       eprint = {astro-ph/0612031},
 primaryClass = {astro-ph},
       adsurl = {https://ui.adsabs.harvard.edu/abs/2007ApJ...662.1093W}
}

@ARTICLE{2006ApJ...642..354Z,
       author = {{Zhang}, Bing and {Fan}, Y.~Z. and {Dyks}, Jaroslaw and {Kobayashi}, Shiho and {M{\'e}sz{\'a}ros}, Peter and {Burrows}, David N. and {Nousek}, John A. and {Gehrels}, Neil},
        title = "{Physical Processes Shaping Gamma-Ray Burst X-Ray Afterglow Light Curves: Theoretical Implications from the Swift X-Ray Telescope Observations}",
      journal = {\apj},
         year = 2006,
        month = may,
       volume = {642},
       number = {1},
        pages = {354-370},
          doi = {10.1086/500723},
archivePrefix = {arXiv},
       eprint = {astro-ph/0508321},
 primaryClass = {astro-ph},
       adsurl = {https://ui.adsabs.harvard.edu/abs/2006ApJ...642..354Z}
}

@ARTICLE{2020MNRAS.494.5259Y,
       author = {{Yamazaki}, Ryo and {Sato}, Yuri and {Sakamoto}, Takanori and {Serino}, Motoko},
        title = "{Less noticeable shallow decay phase in early X-ray afterglows of GeV/TeV-detected gamma-ray bursts}",
      journal = {\mnras},
         year = 2020,
        month = jun,
       volume = {494},
       number = {4},
        pages = {5259-5269},
          doi = {10.1093/mnras/staa1095},
archivePrefix = {arXiv},
       eprint = {1910.04097},
 primaryClass = {astro-ph.HE},
       adsurl = {https://ui.adsabs.harvard.edu/abs/2020MNRAS.494.5259Y}
}

@ARTICLE{2025ApJ...985..261D,
       author = {{Dereli-B{\'e}gu{\'e}}, H. and {Pe'er}, A. and {B{\'e}gu{\'e}}, D. and {Ryde}, F.},
        title = "{Unraveling the Origins of Gamma-Ray Burst X-Ray Plateaus through a Study of X-Ray Flares}",
      journal = {\apj},
         year = 2025,
        month = jun,
       volume = {985},
       number = {2},
          eid = {261},
        pages = {261},
          doi = {10.3847/1538-4357/adcead},
archivePrefix = {arXiv},
       eprint = {2412.11533},
 primaryClass = {astro-ph.HE},
       adsurl = {https://ui.adsabs.harvard.edu/abs/2025ApJ...985..261D}
}

@ARTICLE{2016ApJ...822...68A,
       author = {{Ackermann}, M. and {Ajello}, M. and {Anderson}, B. and {Atwood}, W.~B. and {Axelsson}, M. and {Baldini}, L. and {Barbiellini}, G. and {Bastieri}, D. and {Bellazzini}, R. and {Bhat}, P.~N. and {Bissaldi}, E. and {Bonino}, R. and {Bottacini}, E. and {Brandt}, T.~J. and {Bregeon}, J. and {Bruel}, P. and {Buehler}, R. and {Buson}, S. and {Caliandro}, G.~A. and {Cameron}, R.~A. and {Caragiulo}, M. and {Caraveo}, P.~A. and {Cecchi}, C. and {Charles}, E. and {Chekhtman}, A. and {Chiang}, J. and {Chiaro}, G. and {Ciprini}, S. and {Claus}, R. and {Cohen-Tanugi}, J. and {Conrad}, J. and {Cutini}, S. and {D'Ammando}, F. and {de Angelis}, A. and {de Palma}, F. and {Desiante}, R. and {Di Venere}, L. and {Drell}, P.~S. and {Favuzzi}, C. and {Focke}, W.~B. and {Franckowiak}, A. and {Funk}, S. and {Fusco}, P. and {Gargano}, F. and {Gasparrini}, D. and {Gehrels}, N. and {Giglietto}, N. and {Giordano}, F. and {Giroletti}, M. and {Godfrey}, G. and {Grenier}, I.~A. and {Grove}, J.~E. and {Guiriec}, S. and {Hewitt}, J.~W. and {Hill}, A.~B. and {Horan}, D. and {J{\'o}hannesson}, G. and {Kocevski}, D. and {Kouveliotou}, C. and {Kuss}, M. and {Larsson}, S. and {Li}, J. and {Li}, L. and {Longo}, F. and {Loparco}, F. and {Lovellette}, M.~N. and {Lubrano}, P. and {Mayer}, M. and {Mazziotta}, M.~N. and {McEnery}, J.~E. and {Michelson}, P.~F. and {Mizuno}, T. and {Monzani}, M.~E. and {Morselli}, A. and {Murgia}, S. and {Nemmen}, R. and {Nuss}, E. and {Ohno}, M. and {Ohsugi}, T. and {Omodei}, N. and {Orienti}, M. and {Orlando}, E. and {Paneque}, D. and {Perkins}, J.~S. and {Pesce-Rollins}, M. and {Piron}, F. and {Pivato}, G. and {Porter}, T.~A. and {Racusin}, J.~L. and {Rain{\`o}}, S. and {Rando}, R. and {Razzano}, M. and {Reimer}, A. and {Reimer}, O. and {Schaal}, M. and {Schulz}, A. and {Sgr{\`o}}, C. and {Siskind}, E.~J. and {Spada}, F. and {Spandre}, G. and {Spinelli}, P. and {Takahashi}, H. and {Thayer}, J.~B. and {Tibaldo}, L. and {Tinivella}, M. and {Torres}, D.~F. and {Tosti}, G. and {Troja}, E. and {Vianello}, G. and {von Kienlin}, A. and {Werner}, M. and {Wood}, K.~S.},
        title = "{Fermi LAT Stacking Analysis of Swift Localized GRBs}",
      journal = {\apj},
         year = 2016,
        month = may,
       volume = {822},
       number = {2},
          eid = {68},
        pages = {68},
          doi = {10.3847/0004-637X/822/2/68},
archivePrefix = {arXiv},
       eprint = {1605.02096},
 primaryClass = {astro-ph.HE},
       adsurl = {https://ui.adsabs.harvard.edu/abs/2016ApJ...822...68A}
}

@ARTICLE{2022NatCo..13.5611D,
       author = {{Dereli-B{\'e}gu{\'e}}, H{\"u}sne and {Pe'er}, Asaf and {Ryde}, Felix and {Oates}, Samantha R. and {Zhang}, Bing and {Dainotti}, Maria G.},
        title = "{A wind environment and Lorentz factors of tens explain gamma-ray bursts X-ray plateau}",
      journal = {Nat. Commun.},
         year = 2022,
        month = sep,
       volume = {13},
          eid = {5611},
        pages = {5611},
          doi = {10.1038/s41467-022-32881-1},
archivePrefix = {arXiv},
       eprint = {2207.11066},
 primaryClass = {astro-ph.HE},
       adsurl = {https://ui.adsabs.harvard.edu/abs/2022NatCo..13.5611D}
}

@ARTICLE{2020MNRAS.492.2847B,
       author = {{Beniamini}, Paz and {Duque}, Rapha{\"e}l and {Daigne}, Fr{\'e}d{\'e}ric and {Mochkovitch}, Robert},
        title = "{X-ray plateaus in gamma-ray bursts' light curves from jets viewed slightly off-axis}",
      journal = {\mnras},
         year = 2020,
        month = feb,
       volume = {492},
       number = {2},
        pages = {2847-2857},
          doi = {10.1093/mnras/staa070},
archivePrefix = {arXiv},
       eprint = {1907.05899},
 primaryClass = {astro-ph.HE},
       adsurl = {https://ui.adsabs.harvard.edu/abs/2020MNRAS.492.2847B}
}

@ARTICLE{2024MNRAS.528.4307B,
       author = {{Bo{\v{s}}njak}, {\v{Z}}eljka and {Zhang}, B. Theodore and {Murase}, Kohta and {Ioka}, Kunihito},
        title = "{Off-axis MeV and very-high-energy gamma-ray emissions from structured gamma-ray burst jets}",
      journal = {\mnras},
         year = 2024,
        month = mar,
       volume = {528},
       number = {3},
        pages = {4307-4313},
          doi = {10.1093/mnras/stae093},
archivePrefix = {arXiv},
       eprint = {2306.14729},
 primaryClass = {astro-ph.HE},
       adsurl = {https://ui.adsabs.harvard.edu/abs/2024MNRAS.528.4307B}
}

@ARTICLE{2013MNRAS.428..729M,
       author = {{Margutti}, R. and {Zaninoni}, E. and {Bernardini}, M.~G. and {Chincarini}, G. and {Pasotti}, F. and {Guidorzi}, C. and {Angelini}, L. and {Burrows}, D.~N. and {Capalbi}, M. and {Evans}, P.~A. and {Gehrels}, N. and {Kennea}, J. and {Mangano}, V. and {Moretti}, A. and {Nousek}, J. and {Osborne}, J.~P. and {Page}, K.~L. and {Perri}, M. and {Racusin}, J. and {Romano}, P. and {Sbarufatti}, B. and {Stafford}, S. and {Stamatikos}, M.},
        title = "{The prompt-afterglow connection in gamma-ray bursts: a comprehensive statistical analysis of Swift X-ray light curves}",
      journal = {\mnras},
         year = 2013,
        month = jan,
       volume = {428},
       number = {1},
        pages = {729-742},
          doi = {10.1093/mnras/sts066},
archivePrefix = {arXiv},
       eprint = {1203.1059},
 primaryClass = {astro-ph.HE},
       adsurl = {https://ui.adsabs.harvard.edu/abs/2013MNRAS.428..729M}
}

@ARTICLE{2006ApJ...642..389N,
       author = {{Nousek}, J.~A. and {Kouveliotou}, C. and {Grupe}, D. and {Page}, K.~L. and {Granot}, J. and {Ramirez-Ruiz}, E. and {Patel}, S.~K. and {Burrows}, D.~N. and {Mangano}, V. and {Barthelmy}, S. and {Beardmore}, A.~P. and {Campana}, S. and {Capalbi}, M. and {Chincarini}, G. and {Cusumano}, G. and {Falcone}, A.~D. and {Gehrels}, N. and {Giommi}, P. and {Goad}, M.~R. and {Godet}, O. and {Hurkett}, C.~P. and {Kennea}, J.~A. and {Moretti}, A. and {O'Brien}, P.~T. and {Osborne}, J.~P. and {Romano}, P. and {Tagliaferri}, G. and {Wells}, A.~A.},
        title = "{Evidence for a Canonical Gamma-Ray Burst Afterglow Light Curve in the Swift XRT Data}",
      journal = {\apj},
         year = 2006,
        month = may,
       volume = {642},
       number = {1},
        pages = {389-400},
          doi = {10.1086/500724},
archivePrefix = {arXiv},
       eprint = {astro-ph/0508332},
 primaryClass = {astro-ph},
       adsurl = {https://ui.adsabs.harvard.edu/abs/2006ApJ...642..389N}
}

@ARTICLE{2006ApJ...647.1213O,
       author = {{O'Brien}, P.~T. and {Willingale}, R. and {Osborne}, J. and {Goad}, M.~R. and {Page}, K.~L. and {Vaughan}, S. and {Rol}, E. and {Beardmore}, A. and {Godet}, O. and {Hurkett}, C.~P. and {Wells}, A. and {Zhang}, B. and {Kobayashi}, S. and {Burrows}, D.~N. and {Nousek}, J.~A. and {Kennea}, J.~A. and {Falcone}, A. and {Grupe}, D. and {Gehrels}, N. and {Barthelmy}, S. and {Cannizzo}, J. and {Cummings}, J. and {Hill}, J.~E. and {Krimm}, H. and {Chincarini}, G. and {Tagliaferri}, G. and {Campana}, S. and {Moretti}, A. and {Giommi}, P. and {Perri}, M. and {Mangano}, V. and {LaParola}, V.},
        title = "{The Early X-Ray Emission from GRBs}",
      journal = {\apj},
         year = 2006,
        month = aug,
       volume = {647},
       number = {2},
        pages = {1213-1237},
          doi = {10.1086/505457},
archivePrefix = {arXiv},
       eprint = {astro-ph/0601125},
 primaryClass = {astro-ph},
       adsurl = {https://ui.adsabs.harvard.edu/abs/2006ApJ...647.1213O}
}

@book{Agresti2002,
author = {Agresti, A.},
title= {Categorical Data Analysis (Second Edition)},
year = {2002},
doi = {10.1002/0471249688},
address = {Hoboken, New Jersey},
publisher = {John Wiley \& Sons},
}

@article{54475ba4-6108-364d-b047-e5384517b947,
 ISSN = {0006341X, 15410420},
 URL = {http://www.jstor.org/stable/3001616},
 author = {William G. Cochran},
 journal = {Biom.},
 number = {4},
 pages = {417--451},
 publisher = {[Wiley, International Biometric Society]},
 title = {Some Methods for Strengthening the Common Chi-Square Tests},
 urldate = {2025-10-16},
 volume = {10},
 year = {1954}
}

@article{10.1093/jnci/22.4.719,
    author = {Mantel, Nathan and Haenszel, William},
    title = {Statistical Aspects of the Analysis of Data From Retrospective Studies of Disease},
    journal = {JNCI},
    volume = {22},
    number = {4},
    pages = {719-748},
    year = {1959},
    month = {04},
    issn = {0027-8874},
    doi = {10.1093/jnci/22.4.719},
    url = {https://doi.org/10.1093/jnci/22.4.719},
    eprint = {https://academic.oup.com/jnci/article-pdf/22/4/719/2674652/22-4-719.pdf},
}

@book{mccullagh2019generalized,
  title={Generalized linear models},
  author={McCullagh, Peter},
  year={2019},
  publisher={Routledge}
}

@article{497e1044-d5b0-30a9-b230-3ca0f10d6f6c,
 ISSN = {00034851},
 URL = {http://www.jstor.org/stable/2236101},
 author = {H. B. Mann and D. R. Whitney},
 journal = {Ann. Math. Statist.},
 number = {1},
 pages = {50--60},
 publisher = {Institute of Mathematical Statistics},
 title = {On a Test of Whether one of Two Random Variables is Stochastically Larger than the Other},
 urldate = {2025-10-16},
 volume = {18},
 year = {1947}
}

@ARTICLE{2019ApJ...884...59L,
       author = {{Liu}, Chuanxi and {Mao}, Jirong},
        title = "{GRB X-Ray Flare Properties among Different GRB Subclasses}",
      journal = {\apj},
         year = 2019,
        month = oct,
       volume = {884},
       number = {1},
          eid = {59},
        pages = {59},
          doi = {10.3847/1538-4357/ab3e75},
archivePrefix = {arXiv},
       eprint = {1910.04512},
 primaryClass = {astro-ph.HE},
       adsurl = {https://ui.adsabs.harvard.edu/abs/2019ApJ...884...59L}
}

@ARTICLE{2024ApJ...970..141A,
       author = {{Asano}, Katsuaki},
        title = "{Multiwavelength Modeling for the Shallow Decay Phase of Gamma-Ray Burst Afterglows}",
      journal = {\apj},
         year = 2024,
        month = aug,
       volume = {970},
       number = {2},
          eid = {141},
        pages = {141},
          doi = {10.3847/1538-4357/ad6148},
archivePrefix = {arXiv},
       eprint = {2404.09675},
 primaryClass = {astro-ph.HE},
       adsurl = {https://ui.adsabs.harvard.edu/abs/2024ApJ...970..141A}
}

@ARTICLE{2011ApJ...732...77M,
       author = {{Murase}, Kohta and {Toma}, Kenji and {Yamazaki}, Ryo and {M{\'e}sz{\'a}ros}, Peter},
        title = "{On the Implications of Late Internal Dissipation for Shallow-decay Afterglow Emission and Associated High-energy Gamma-ray Signals}",
      journal = {\apj},
         year = 2011,
        month = may,
       volume = {732},
       number = {2},
          eid = {77},
        pages = {77},
          doi = {10.1088/0004-637X/732/2/77},
archivePrefix = {arXiv},
       eprint = {1011.0988},
 primaryClass = {astro-ph.HE},
       adsurl = {https://ui.adsabs.harvard.edu/abs/2011ApJ...732...77M}
}

@ARTICLE{2007A&A...469..379E,
       author = {{Evans}, P.~A. and {Beardmore}, A.~P. and {Page}, K.~L. and {Tyler}, L.~G. and {Osborne}, J.~P. and {Goad}, M.~R. and {O'Brien}, P.~T. and {Vetere}, L. and {Racusin}, J. and {Morris}, D. and {Burrows}, D.~N. and {Capalbi}, M. and {Perri}, M. and {Gehrels}, N. and {Romano}, P.},
        title = "{An online repository of Swift/XRT light curves of {\ensuremath{\gamma}}-ray bursts}",
      journal = {\aap},
         year = 2007,
        month = jul,
       volume = {469},
       number = {1},
        pages = {379-385},
          doi = {10.1051/0004-6361:20077530},
archivePrefix = {arXiv},
       eprint = {0704.0128},
 primaryClass = {astro-ph},
       adsurl = {https://ui.adsabs.harvard.edu/abs/2007A&A...469..379E}
}

@ARTICLE{2019ApJ...878...52A,
       author = {{Ajello}, M. and {Arimoto}, M. and {Axelsson}, M. and {Baldini}, L. and {Barbiellini}, G. and {Bastieri}, D. and {Bellazzini}, R. and {Bhat}, P.~N. and {Bissaldi}, E. and {Blandford}, R.~D. and {Bonino}, R. and {Bonnell}, J. and {Bottacini}, E. and {Bregeon}, J. and {Bruel}, P. and {Buehler}, R. and {Cameron}, R.~A. and {Caputo}, R. and {Caraveo}, P.~A. and {Cavazzuti}, E. and {Chen}, S. and {Cheung}, C.~C. and {Chiaro}, G. and {Ciprini}, S. and {Costantin}, D. and {Crnogorcevic}, M. and {Cutini}, S. and {Dainotti}, M. and {D'Ammando}, F. and {de la Torre Luque}, P. and {de Palma}, F. and {Desai}, A. and {Desiante}, R. and {Di Lalla}, N. and {Di Venere}, L. and {Fana Dirirsa}, F. and {Fegan}, S.~J. and {Franckowiak}, A. and {Fukazawa}, Y. and {Funk}, S. and {Fusco}, P. and {Gargano}, F. and {Gasparrini}, D. and {Giglietto}, N. and {Giordano}, F. and {Giroletti}, M. and {Green}, D. and {Grenier}, I.~A. and {Grove}, J.~E. and {Guiriec}, S. and {Hays}, E. and {Hewitt}, J.~W. and {Horan}, D. and {J{\'o}hannesson}, G. and {Kocevski}, D. and {Kuss}, M. and {Latronico}, L. and {Li}, J. and {Longo}, F. and {Loparco}, F. and {Lovellette}, M.~N. and {Lubrano}, P. and {Maldera}, S. and {Manfreda}, A. and {Mart{\'\i}-Devesa}, G. and {Mazziotta}, M.~N. and {Mereu}, I. and {Meyer}, M. and {Michelson}, P.~F. and {Mirabal}, N. and {Mitthumsiri}, W. and {Mizuno}, T. and {Monzani}, M.~E. and {Moretti}, E. and {Morselli}, A. and {Moskalenko}, I.~V. and {Negro}, M. and {Nuss}, E. and {Ohno}, M. and {Omodei}, N. and {Orienti}, M. and {Orlando}, E. and {Palatiello}, M. and {Paliya}, V.~S. and {Paneque}, D. and {Persic}, M. and {Pesce-Rollins}, M. and {Petrosian}, V. and {Piron}, F. and {Poolakkil}, S. and {Poon}, H. and {Porter}, T.~A. and {Principe}, G. and {Racusin}, J.~L. and {Rain{\`o}}, S. and {Rando}, R. and {Razzano}, M. and {Razzaque}, S. and {Reimer}, A. and {Reimer}, O. and {Reposeur}, T. and {Ryde}, F. and {Serini}, D. and {Sgr{\`o}}, C. and {Siskind}, E.~J. and {Sonbas}, E. and {Spandre}, G. and {Spinelli}, P. and {Suson}, D.~J. and {Tajima}, H. and {Takahashi}, M. and {Tak}, D. and {Thayer}, J.~B. and {Torres}, D.~F. and {Troja}, E. and {Valverde}, J. and {Veres}, P. and {Vianello}, G. and {von Kienlin}, A. and {Wood}, K. and {Yassine}, M. and {Zhu}, S. and {Zimmer}, S.},
        title = "{A Decade of Gamma-Ray Bursts Observed by Fermi-LAT: The Second GRB Catalog}",
      journal = {\apj},
         year = 2019,
        month = jun,
       volume = {878},
       number = {1},
          eid = {52},
        pages = {52},
          doi = {10.3847/1538-4357/ab1d4e},
archivePrefix = {arXiv},
       eprint = {1906.11403},
 primaryClass = {astro-ph.HE},
       adsurl = {https://ui.adsabs.harvard.edu/abs/2019ApJ...878...52A}
}

@article {Bland170,
	author = {Bland, J Martin and Altman, Douglas G},
	title = {Multiple significance tests: the Bonferroni method},
	volume = {310},
	number = {6973},
	pages = {170},
	year = {1995},
	doi = {10.1136/bmj.310.6973.170},
	publisher = {BMJ Publishing Group Ltd},
	issn = {0959-8138},
	URL = {https://www.bmj.com/content/310/6973/170},
	eprint = {https://www.bmj.com/content/310/6973/170.full.pdf},
	journal = {BMJ}
}

@book{dodge2008concise,
  address = {New York },
  author = {Dodge, Yadolah},
  edition = {First},
  publisher = {Springer },
  title = {The Concise Encyclopedia of Statistics },
  year = 2008
}

@ARTICLE{2006MNRAS.369..311Y,
       author = {{Yamazaki}, Ryo and {Toma}, Kenji and {Ioka}, Kunihito and {Nakamura}, Takashi},
        title = "{Tail emission of prompt gamma-ray burst jets}",
      journal = {\mnras},
         year = 2006,
        month = jun,
       volume = {369},
       number = {1},
        pages = {311-316},
          doi = {10.1111/j.1365-2966.2006.10290.x},
archivePrefix = {arXiv},
       eprint = {astro-ph/0509159},
 primaryClass = {astro-ph},
       adsurl = {https://ui.adsabs.harvard.edu/abs/2006MNRAS.369..311Y}
}

\begin{appendix}

\onecolumn
\section{Tables}
\begin{table}[h!]
\caption{\label{tab:complexity_bins}Stratified comparison of afterglow complexity}
\centering
\begin{tabular}{lccccccc}
\hline
$t_{XRT}$ bin [s] & $N_{\mathrm{NoHE}}$ & $N_{\mathrm{HE}}$ & $\chi^{2}$ & $p_{\chi^{2}}$ & $U$ & $p_{\mathrm{MW}}$ \\
\hline
30 – 300        & 908 & 38 & 8.21 & 0.14  & 13828.5 & 0.033 \\
300 – 3000      & 79   & 6  & 0.12 & 0.94  & 240.0   & 0.96  \\
3000 – 30000    & 99  & 12 & 3.52 & 0.17  & 431.0   & 0.08 \\
30000 – 300000  & 36   & 36 & 3.60 & 0.16  & 539.5   & 0.06  \\
\hline
\end{tabular}
\tablefoot{\centering Comparison of the number of breakpoints (\(n_{\mathrm{breaks}}\)) between GRBs with and without high-energy (HE) detection, stratified by data-taking starting time $t_{XRT}$. For each bin, the number of bursts in the two groups is reported together with chi-square and Mann–Whitney U test statistics. Although the earliest bin (30–300 s) yields a nominal Mann–Whitney p value of 0.033, no bin remains statistically significant after correction for multiple comparisons (Bonferroni adjustment over m=8 bin-wise tests).}
\end{table}

\begin{table}[h!]
\caption{\label{tab:GLM_complexity}Poisson GLM results for afterglow complexity}
\centering
\begin{tabular}{lcccc}
\hline
Coefficient & Estimate & Std. Error & $z=\frac{\text{Coefficient}}{\text{Std. Error}}$ & $p$-value \\
\hline
$\beta_0$ & 2.29 & 0.109 & 21.09 & $<0.0001$ \\
$\beta_{HE}$ & $-0.16$ & 0.25 & $-0.67$ & 0.5 \\
$\beta_{t_{XRT}}$ & $-0.77$ & 0.053 & $-14.46$ & $<0.0001$ \\
$\beta_{int}$ & 0.14 & 0.11 & 1.26 & 0.21 \\
\hline
\multicolumn{5}{l}{Model summary: Log-likelihood = $-1871.4$, Deviance = 1186.2, Pearson $\chi^2 = 1050$,} \\
\multicolumn{5}{c}{df = 1211, pseudo-$R^2_{\mathrm{CS}}=0.28$, overdispersion ratio = 0.87.} \\
\hline
\end{tabular}
\tablefoot{\centering Results of the Poisson generalized linear model for the number of breaks ($n_{\mathrm{breaks}}$). HC3 standard errors are reported.}
\end{table}

\begin{table}[h!]
\caption{\label{tab:plateau_bins}Plateau incidence in stratified HE and non-HE samples}
\centering
\begin{tabular}{lcccccccc}
\hline
$t_{XRT}$ bin [s] & $N_{\mathrm{NoHE-NoPlat}}$ & $N_{\mathrm{NoHE-Plat}}$ & $N_{\mathrm{HE-NoPlat}}$ & $N_{\mathrm{HE-Plat}}$ & OR (Fisher) & CI low & CI high & $p_{\mathrm{Fisher}}$ \\
\hline
30 – 300        & 618 & 290 & 26 & 12 & 0.98 & 0.49 & 1.98 & 1.0\\
300 – 3000      & 66   & 13  & 6 & 0 & 0.38 & 0.020 & 7.13 & 0.58 \\
3000 – 30000    & 98  & 1 & 12 & 0 & 2.63 & 0.10 & 68.03 & 1.0 \\
30000 – 300000  & 36   & 0 & 33 & 3 & 7.63 & 0.38 & 153.20 & 0.24 \\
pooled (CMH) &  -- & -- & -- & -- & 1.04 & 0.54 & 1.98 & 0.91 \\
\hline
\end{tabular}
\tablefoot{\centering Comparison of plateau incidence between GRBs with and without HE detection, stratified by the XRT observation start time $t_{XRT}$.
Odds ratios (OR) are derived from Fisher’s exact test for each bin and for the pooled sample.}
\end{table}

\begin{table}[h!]
\caption{\label{tab:GLM_plat}Binomial GLM results for plateau incidence}
\centering
\begin{tabular}{lcccc}
\hline
Coefficient & Estimate & Std. Error & $z=\frac{\text{Coefficient}}{\text{Std. Error}}$ & $p$-value \\
\hline
$\alpha_0$ & 1.79 & 0.38 & 4.75 & $<0.0001$ \\
$\alpha_{HE}$ & $-0.0016$ & 0.33 & $-0.005$ & 1.0 \\
$\alpha_{t_{XRT}}$ & $-1.29$ & 0.18 & $-7.23$ & $<0.0001$ \\
\hline
\multicolumn{5}{l}{Model summary: Log-likelihood = $-644.89$, Deviance = 1289.8, Pearson $\chi^2 = 1280$,} \\
\multicolumn{5}{c}{df = 1212, pseudo-$R^2_{\mathrm{CS}}=0.086$.} \\
\hline
\end{tabular}
\tablefoot{\centering HC3 standard errors are reported.}
\end{table}

\FloatBarrier
\twocolumn
\section{$\alpha_1$ versus high-energy detection}
\label{sec:a1}
\begin{figure}
\centering
\begin{minipage}{.45\textwidth}
  \centering
  \includegraphics[width=\hsize]{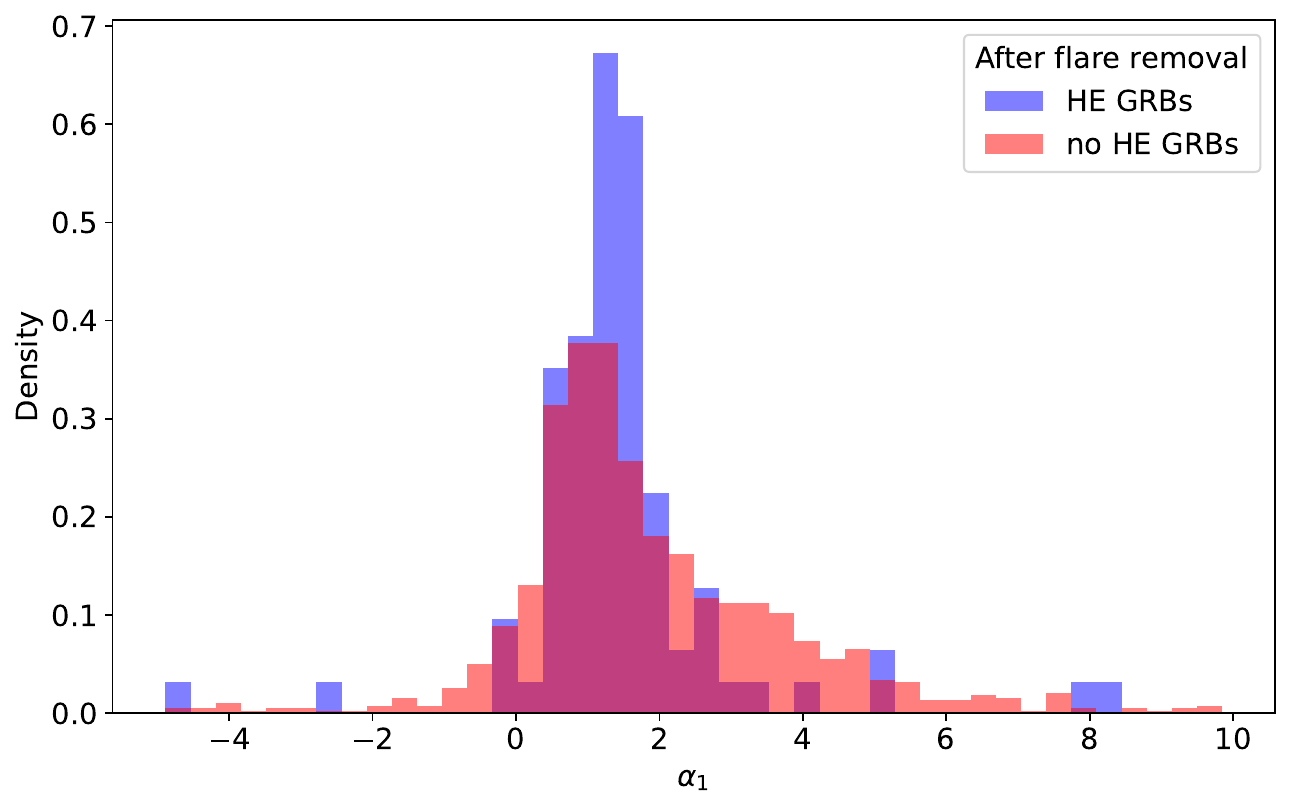}
  \captionof{figure}{Comparison of the distributions of the slope of the first power-law segment $\alpha_1$ between GRBs with (blue) and without (red) a high-energy detection.}
  \label{a1_tot}
\end{minipage}
\begin{minipage}{.45\textwidth}
  \centering
  \includegraphics[width=\hsize]{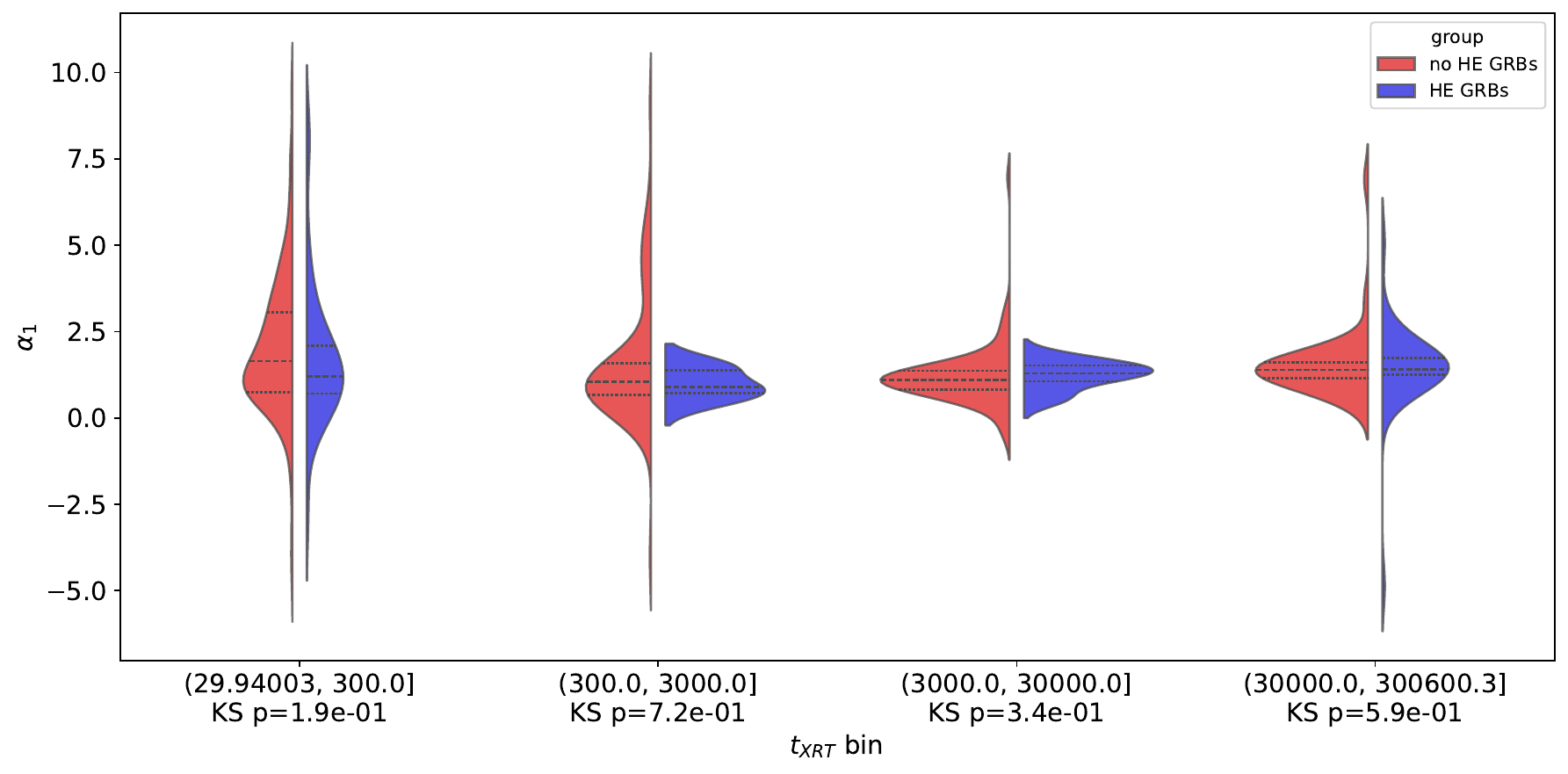}
  \captionof{figure}{Violin plot showing the comparison of the distributions of $\alpha_1$ in the different $t_{XRT}$ bins.}
  \label{a1_t0}
\end{minipage}
\end{figure}
We tested whether there was a difference in the distributions of the slope of the first power-law segments $\alpha_1$ between GRBs with and without a high-energy detection.
Once again, the direct comparison of the two samples (without accounting for $t_{XRT}$) returns two significantly different distributions ($\alpha_1^{peak}=1.35,\,\sigma_{\alpha_1}=6.72$ and $\alpha_1^{peak}=1.19,\,\sigma_{\alpha_1}=3.17$ for GRBs with and without a HE detection respectively), with a Kolmogorov–Smirnov test (KS test, \cite{dodge2008concise}) returning a $p=4.9\times10^{-4}$ (Fig. \ref{a1_tot}).
However, dividing the data into $t_{XRT}$ bins completely removes this difference, with the resulting distributions showing no significant differences (KS p values $\geq10^{-1}$, Fig. \ref{a1_t0}).

\section{Other results}
\label{sec:redshift}
\begin{figure}
\centering
\begin{minipage}{.45\textwidth}
  \centering
  \includegraphics[width=\hsize]{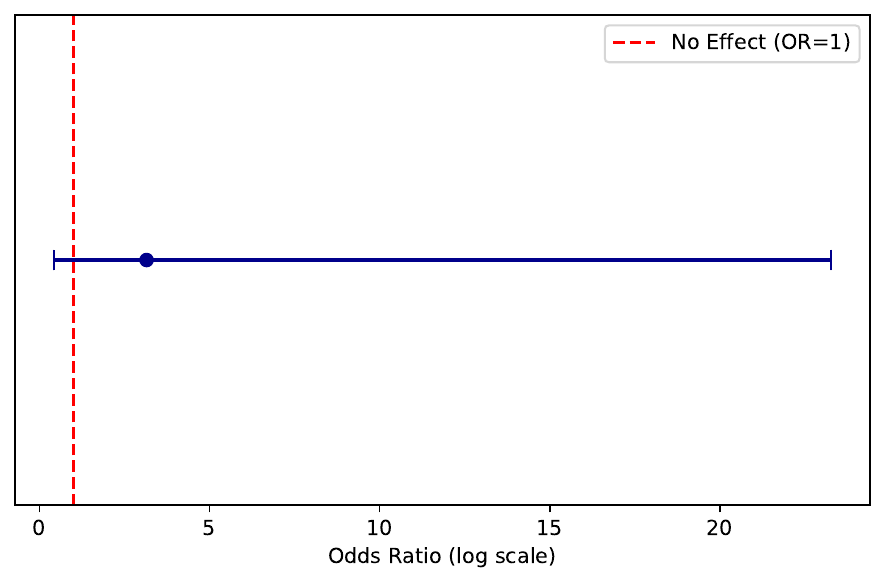}
\captionof{figure}{Forest plot showing the odds ratio (OR) and 95\% confidence interval for the association between HE detection and Long GRBs (OR of Long GRB given a HE detection). The vertical dashed red line marks the null hypothesis of no association $(OR = 1)$.}
\label{fig:LvsS}
\end{minipage}
\begin{minipage}{0.45\textwidth}
  \centering
  \includegraphics[width=\hsize]{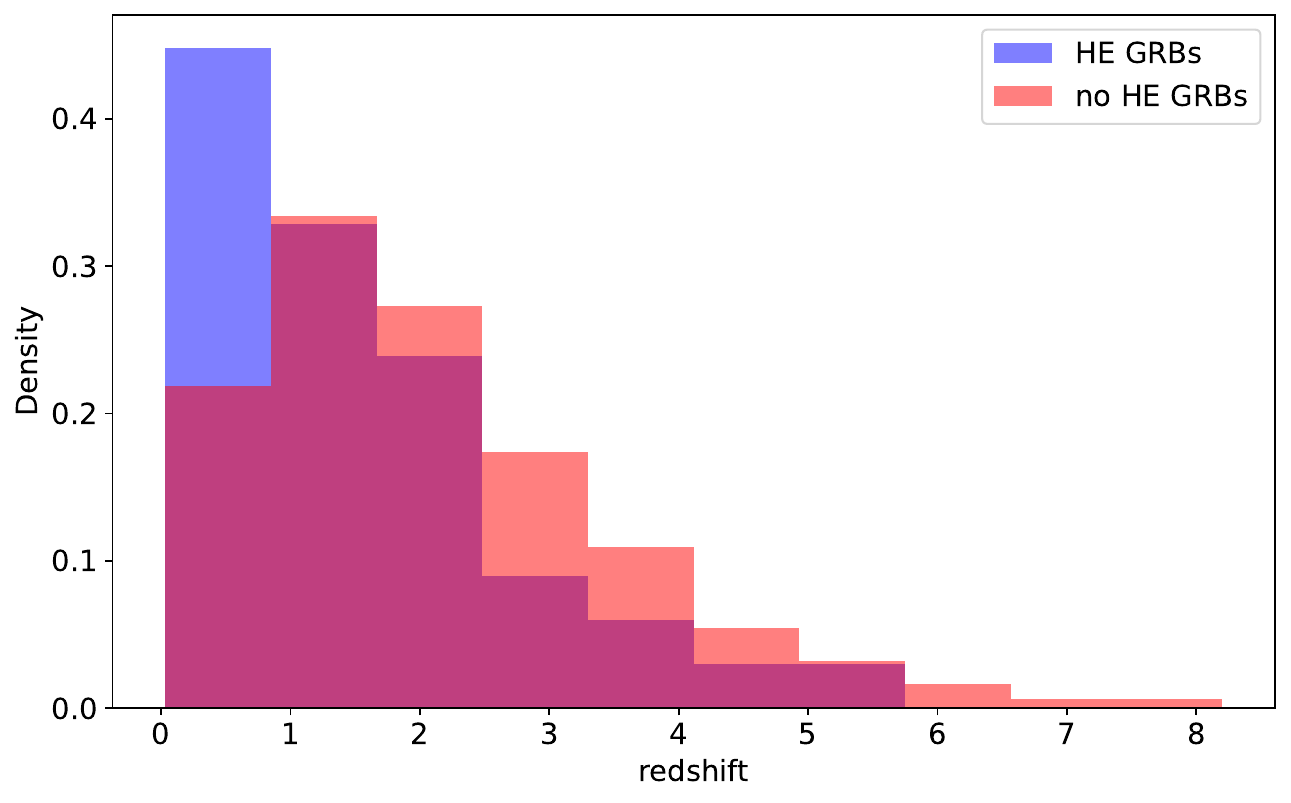}
  \captionof{figure}{Redshift distributions for the GRBs in the two groups.}
  \label{fig:redshift}
\end{minipage}
\end{figure}

The two families of GRBs (with and without a HE/VHE detection) show no significant differences in the fraction of short GRBs compared to long ones.
The group of GRBs with HE/VHE detection is composed of $97.67\%$ long GRBs, whereas in the other group the fraction is $93.02\%$.
The distribution of short vs. long events did not differ significantly between the two groups: $\chi^2$ = 0.77, p = 0.38; Fisher’s exact p = 0.35; odds ratio = 3.15, 95\% CI 0.43-23.27 (Fig. \ref{fig:LvsS}).

Figure \ref{fig:redshift} shows the redshift distributions for those GRBs in the two groups for which this measure is available.
There is an excess of low-redshift GRBs among those seen also in HE compared to the other group and a two-sample KS test returned a statistic $D$ of 0.23 and a $p$-value of $0.036$, indicating a mildly significant difference between the two distributions.

\section{Comparison with the Swift pipeline}
\label{sec:confront}
\begin{figure}
\centering
\begin{minipage}{.45\textwidth}
  \centering
  \includegraphics[width=\hsize]{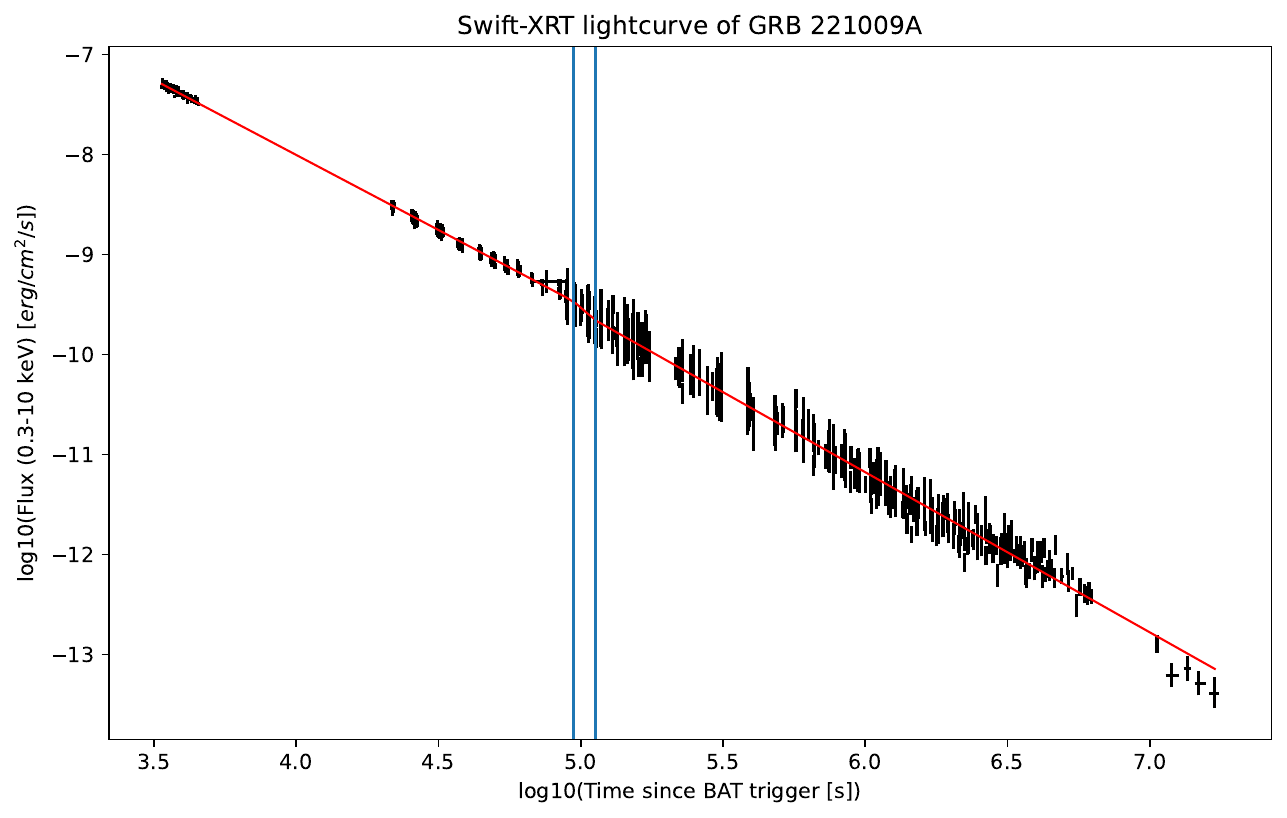}
  \captionof{figure}{GRB 221009A analyzed with the new pipeline: no flare and 2 breakpoints identified (blue vertical lines).}
  \label{BOATnew}
\end{minipage}
\begin{minipage}{.45\textwidth}
  \centering
  \includegraphics[width=\hsize]{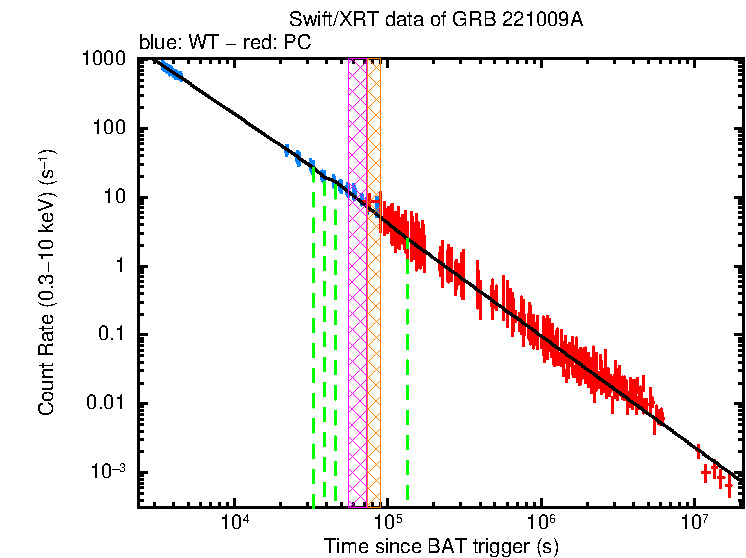}
  \captionof{figure}{GRB 221009A analysis with the Swift automated method: 2 flares (canceled regions) and 4 breakpoints identified (green vertical lines). Credits: \href{https://www.swift.ac.uk/xrt_live_cat/01126853/}{\textit{Swift}-XRT repository of GRBs}}
  \label{BOATSwift}
\end{minipage}
\end{figure}
In this section we report the confrontation between our new analysis pipeline and the official automated \textit{Swift}-XRT analysis \citep{2007A&A...469..379E, 2009MNRAS.397.1177E}.
This was done as a cross-check for the new analysis pipeline.
In figures \ref{BOATnew} and \ref{BOATSwift} the analysis of the BOAT GRB 221009A is reported as an example of the different results obtained with the two different analysis methods: our new analysis pipeline and the official automated \textit{Swift}-XRT analysis \citep{2007A&A...469..379E, 2009MNRAS.397.1177E}.
The official \textit{Swift}-XRT analysis identified two flares between $5.56\times10^4-7.31\times10^4$ s and $7.35\times10^4-8.94\times10^4$ s (the canceled regions in the plot), as well as using four breakpoints to fit the remaining light-curve (vertical green lines, Fig. \ref{BOATSwift}).
Meanwhile, the new analysis pipeline developed here identified no flaring events and used only two breakpoints (blue vertical lines) to best fit the light-curve (Fig. \ref{BOATnew}).
In this case, the simpler model in the current analysis, selected by BIC, seems to be more aligned with the expectations given the canonical light-curve shape to describe the emission structure.

\begin{figure}
\centering
\begin{minipage}{.45\textwidth}
  \centering
  \includegraphics[width=\hsize]{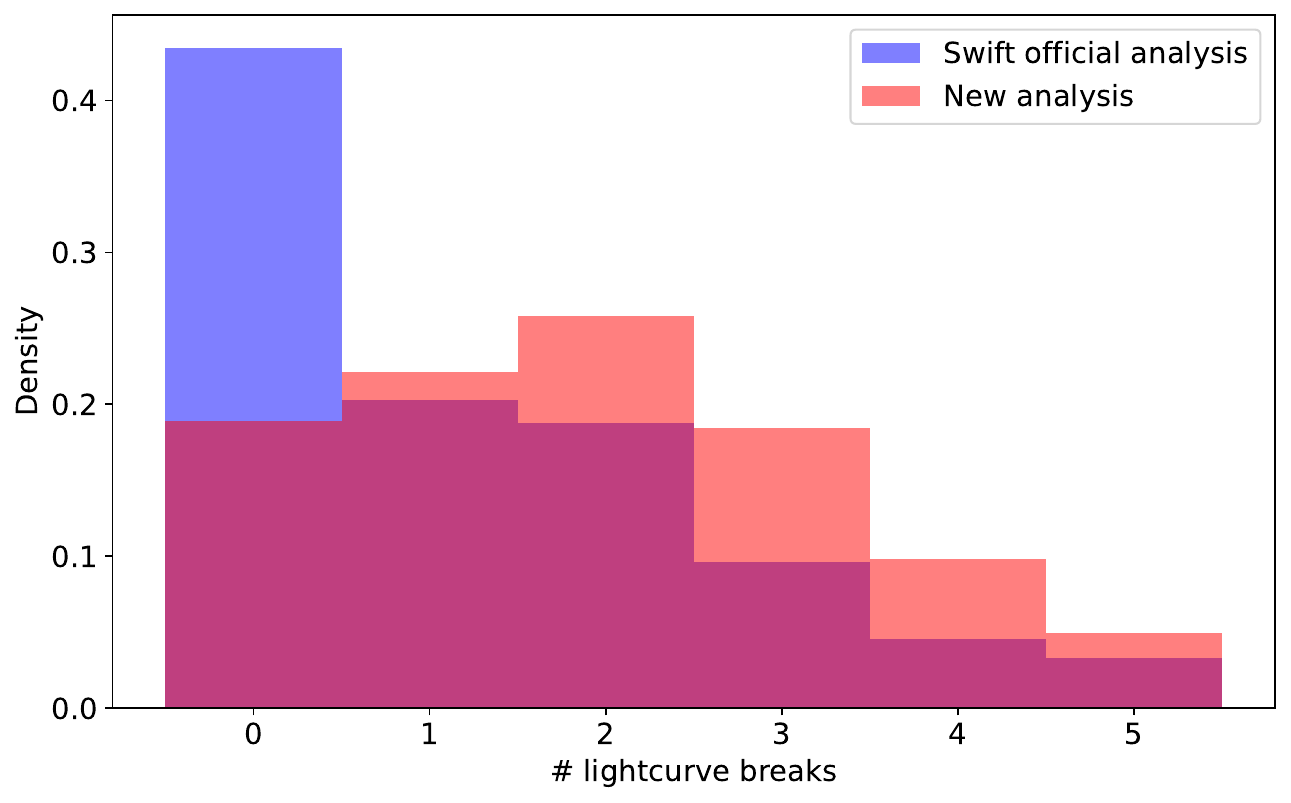}
  \captionof{figure}{Comparison of the distributions of the number of breakpoints used to fit the GRBs in the \textit{Swift}-XRT catalog between the official automated Swift analysis (in blue) and the new analysis pipeline developed in this paper (in red).}
  \label{NvsSbreaks}
\end{minipage}
\begin{minipage}{.45\textwidth}
  \centering
  \includegraphics[width=\hsize]{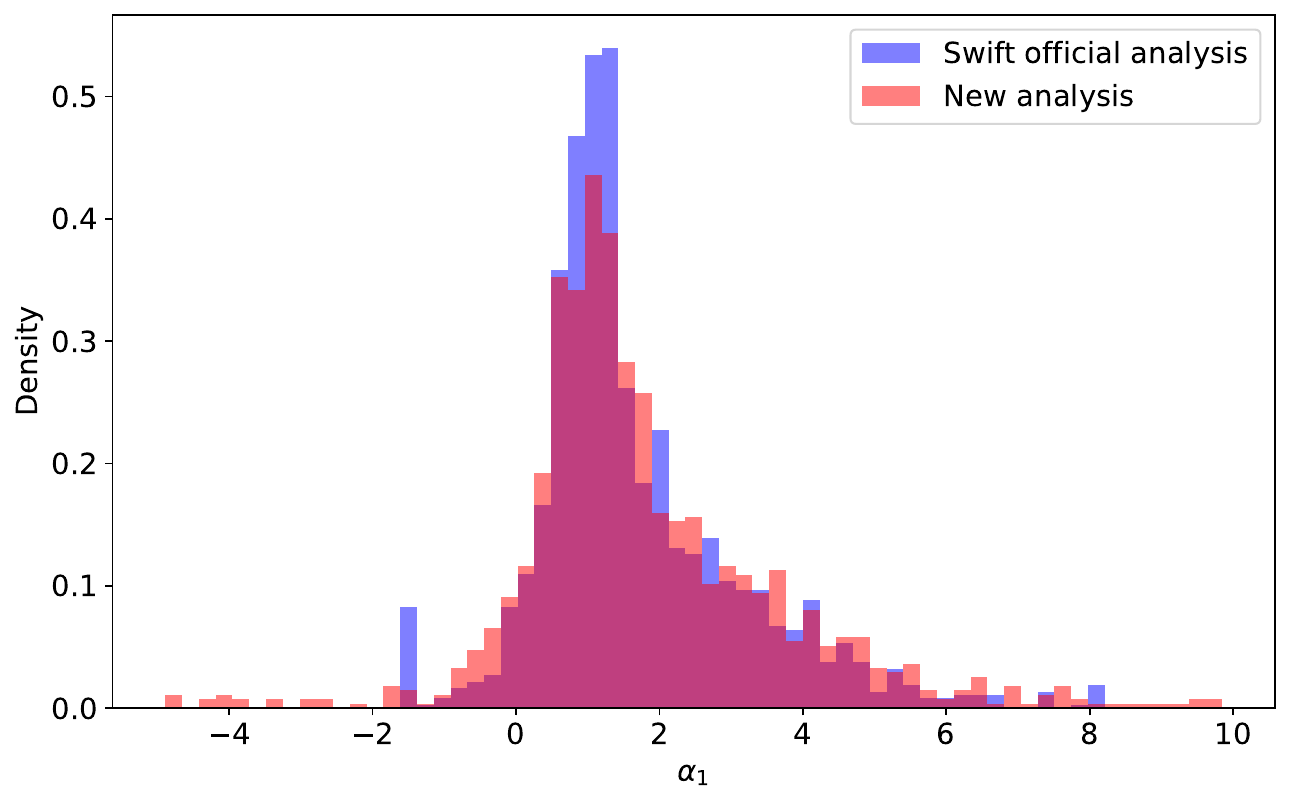}
  \captionof{figure}{Comparison of the distributions of the slope of the first power-law segment of the fit for the GRBs in the \textit{Swift}-XRT catalog between the official automated Swift analysis (in blue) and the new analysis pipeline developed in this paper (in red).}
  \label{NvsSA}
\end{minipage}
\end{figure}

In figures \ref{NvsSbreaks} and \ref{NvsSA} are reported the results of the new analysis carried out on 1438 GRBs reported in the \textit{Swift}-XRT GRB catalog \footnote{\url{https://www.swift.ac.uk/xrt_curves/}} \citep{2007A&A...469..379E, 2009MNRAS.397.1177E}, as well as a comparison with the official automatic Swift analysis.
The histogram in figure \ref{NvsSbreaks} compares the distribution of the number of light-curve breaks identified in GRBs by the Swift official analysis and the new analysis method, while in figure \ref{NvsSA} the histogram compares the distribution of the first power-law slope.

The two approaches differ primarily in the treatment of flares and the statistical criteria used for model selection.
While the Swift method relies on pairwise $\chi^2$ comparisons between models with increasing numbers of breakpoints, the new analysis applies a global comparison using the BIC and incorporates a different, model independent, flare identification procedure.

As a result, the Swift method favors simpler models and shows a peak at zero breaks, whereas the new method peaks at two breaks, aligning more closely with the expectations given the canonical structure for standard X-ray GRB afterglow light-curves \citep{2006ApJ...642..389N, 2006ApJ...642..354Z, 2007ApJ...662.1093W}.

In figure \ref{NvsSA}, the two distributions of the first power-law slope $\alpha_1$ are more similar: both peak near zero and span a comparable range of values, with only slight differences in width and symmetry ($\alpha_1^{peak}=1.03,\,\sigma_{\alpha_1}=1.48$ and $\alpha_1^{peak}=1.23,\,\sigma_{\alpha_1}=3.57$ for GRBs analyzed respectively with the Swift automatic pipeline and with the pipeline outlined in this paper).
Although both distributions are in general agreement with the general findings in the literature (e.g. \cite{2006ApJ...642..389N, 2006MNRAS.369..311Y}), they still show some minor differences.
The Swift analysis yields a slightly narrower distribution with a modest positive skew, whereas the new method produces a slightly broader profile.
This is likely due to the different methods for identification flares that are more abundant in the early phases of the emission.
A KS test on the two $\alpha_1$ distributions gives a statistic $D$ of $7.1\times10^{-2}$ and a $p$-value of $1.89\times10^{-3}$.
Although the small $p$-value suggest the two distributions being significantly different, the modest $D$ statistic, especially in a sample of more than 1400 GRBs, indicates that the practical difference between the two distributions is minor.

\FloatBarrier
\clearpage

\end{appendix}
\end{document}